\documentclass[preprintnumbers,amsmath,amssymb,onecolumn]{revtex4}
\usepackage[dvips,colorlinks=true,citecolor=red,filecolor=green,linkcolor=blue,pdfnewwindow=true]{hyperref}
\usepackage{graphicx}
\usepackage{color}
\usepackage[title]{appendix}

\begin{document}
\title{Network medicine in ovarian cancer: Topological properties to drug discovery}
\author{Keilash Chirom$^{1,2}$, Md. Zubbair Malik$^{1*}$, Pallavi Somvanshi$^{1*}$ and R.K. Brojen Singh$^{1}$}
\email{R.K.B.S.: brojen@jnu.ac.in (Corresponding author),\\
M.Z.M.: zubair.bioinfo@gmail.com (Co-corresponding author), \\
P.S.: psomvanshi@gmail.com (Co-corresponding author)}
\affiliation{$^1$School of Computational $\&$ Integrative Sciences, Jawaharlal Nehru University, New Delhi-110067, India.\\
$^2$Department of Biotechnology, TERI University, New Delhi-110070, India.}

\begin{abstract}
{\noindent}The investigation of topological properties of ovarian cancer network (OCN) and the roles of hubs involved in it by digging the network at various levels of organization are important to understand how OCN is organized to understand disease states. The OCN constructed from the experimentally verified genes exhibits fractal nature in the topological properties of the network and deeply rooted communities. Also, the network properties at all levels of organization obey one parameter scaling law which lacks centrality lethality rule. We then showed that $\langle k\rangle$ can be taken as a scaling parameter, where, power law exponent can be estimated from the ratio of network diameters, $\lambda=1+\rho\left[\frac{d_c}{d_k}\right]$. The betweenness centrality shows two distinct behaviors one shown by high degree hubs and the other by segregated low degree nodes. The $C_B$ distribution follows power law behavior with the exponent connected to exponents of distributions of high and low degree nodes by, $\theta\sim\left[1+\frac{1}{2}\left(\frac{1}{\epsilon_s}+\frac{1}{\epsilon_h}\right)(\lambda-1)\right]$. Absence of rich-club formation leads to the missing of a number of attractors in the network causing formation of weakly tied diverse functional modules to keep optimal network efficiency. The hubs knockout experiment shows that provincial hubs take major responsibility to keep network integrity and organization. The identified key regulators are found to be provincial and connector hubs. Further, two key regulators, EPCAM and CD44 are found to be maximally over expressed at various cancer stages (II-IV). They are also positively correlated with immune infiltrates (CD4+ T cells). Finally, few potential drugs are identified related to the key regulators. \\

\end{abstract}


\maketitle

\noindent\textbf{\large Introduction}\\
Cancer is generally believed to be genetic disease \cite{Kolch} and interferes a number of interacting spatio-temporal signaling gene regulatory networks (SGRN), which are responsible for molecular basis communications, cellular and other processes in the whole complex cellular networks \cite{Erwin}. Since the SGRN are composed of basic motif circuits \cite{Milo}, the overall signal processing by them is the integration of the signals by these circuits \cite{Shoval} which may reflect fractal or scale-free properties in the network organization with hierarchical features \cite{Barabasi} given by self-affine process \cite{Calvert} of any property $\mathcal{F}(x)$ of a scale factor $\alpha$: $\frac{\mathcal{F}(\alpha x)}{\mathcal{F}(x)}=\Gamma(\alpha)\sim\alpha^D$, such that, $\Gamma(\alpha_1\alpha_2...\alpha_N)=\Gamma_1(\alpha_1)\Gamma_2(\alpha_2)...\Gamma_N(\alpha_N): \Gamma_k(\alpha_k)\sim\alpha_k^{D_k}$ with $0\le\alpha,\alpha_1,\alpha_2,...,\alpha_N\le 1$, where, $D,D_k$ are self-similarity dimension. Hence, the static and dynamical changes in the SGRN due to changes in signal processing when the cellular system move from normal to disease state are reflected in the topological properties of the network \cite{Boccaletti}. Because the changes in the signal patterns in the network is driven by genetic and epigenetic perturbations which alter significant rewiring of the these signaling networks \cite{Califano,Schramm}. This perturbation may enhance or diminish the functional activity of a gene or group of genes causing significant changes in the signaling patterns leading to variations in the cellular phenotype \cite{Creixell} which may exhibit in the topological properties of the network \cite{Teschendorff}. On the other hand, Kauffman's \textit{cancer attractor theory} \cite{Kauffman,Huang} developed on the basis of his works and earlier works of Max Delbruck \cite{Delbruck}, and Jacob and Monod \cite{Monod}, is another important theory to explain the disruption in signaling mechanism at cancer state \cite{Huang}. The theory claims that the interacting genes in each SGRN spatio-temporally regulate among themselves to reach an \textit{equilibrium state} or \textit{attractor} which corresponds to balancing of their gene expression patterns \cite{Kauffman}. Hence, the state of a complex network having $N$ interacting genes with their expression levels $x_1,x_2,...,x_N$ at any instant of time $'t'$ can be defined as an $N$-dimensional vector $\vec{\Lambda}(t)=[x_1,x_2,...,x_N]^{-1}$ in the state space \cite{Huang}. In this state space, there could be a large number of \textit{attractors}, where, each \textit{attractor} corresponds to an equilibrium state $S^*$ surrounded by non-stationary unstable states \cite{Kauffman,Kauffman1} and encodes specific genetic activities \cite{Huang}. The attractors could be of two types, one may correspond to stable network corresponding to normal mature cell type, and the other may correspond to \textit{cancer attractor} separated by epigenetic barrier which can be lowered by mutation load. Hence, in the state corresponding to \textit{cancer attractor}, the property of the signal processing mechanism in the network will be quite different and may be measured in the topological properties of the network. It has been proposed that signaling entropy \cite{Gomez}, which can be calculated from local signaling entropy of the network \cite{Teschendorff}, could be one potential parameter to characterize the variations in the signaling patterns in the network. However, capturing these altered signaling patterns with network topological properties and their quantification are still debatable and not fully understood yet.\\

{\noindent}Ovarian cancer is known as silent killer because early diagnosis of this cancer type is quite difficult and generally more than 70\% cases are not able to be detected until at significantly advanced stage \cite{Jayson,Carlson}. One reason could be inability to recognize the symptoms of this cancer because the data indicates that the origin of this cancer is generally from fallopian tube and not from ovaries \cite{Mallen}. Hence, this cancer type is one of the most inheritable cancers, the fifth cause of cancer in women and the fourth cause of women death in developed countries \cite{Berns,Flaum}. Further, since genetic instability is the hallmark of cancer disease that causes genetic heterogeneity triggering neoplastic progression \cite{Hanahan}, it becomes quite difficult to which key gene/genes would be targeted for possible cure of cancer diseases \cite{Merlo}. However, identification of key genes of ovarian cancer should be done systematically from various sensitive experimental and theoretical perspectives because those identified key genes are the hopes of possible drug targets in systematic drug design that may lead to cure and prevent from this dreaded disease. There have been few techniques to identify the key regulators, namely, network theory which was used to identify the five novel key genes (AKT1, KRAS, EPCAM, CD44 and MCAM) in ovarian cancer network  \cite{Malik}, new biomarker (CREB1 gene) that was proposed by analyzing various databases with experiments \cite{Chia}, and GWAS (genome-wide association studies) based identification of biomarkers specially for African women \cite{Manichaikul} etc. \\

{\noindent}In a particular network, hubs (high degree nodes), specially key regulators, are the main agents which are involved in many signaling processes, in maintaining topological properties, stability and self-organization of the network \cite{Barabasi,Albert}. For instance, it has been found that a group of important hub genes involved in prostate cancer are significantly reduced their gene expressions and hence quite disturb in their regulatory mechanisms and signal processing \cite{Ribarska}, regulatory mechanism of a group of genes in few pathways (eg glucose metabolism) in Lung cancer are found to be disturbed \cite{Vanhove}, significant dysregulation of signaling hubs of lipid-regulated genes was found in Hepatocellular carcinoma \cite{Lee} etc. If the cancer cause dysregulation or disturbance in the signaling hubs in the complex gene regulatory network, then there should be significant changes in the topological properties of the network. The question is how to capture the changes in the information patterns in the genes and associated networks. In the complex protein-protein interaction (PPI), one possible technique to estimate the information contained in the network is to measure entropy of the network from which information contained in the network can be estimated \cite{Morzy,Zenil}. The entropy of a network $G(V,N)$ with $N$ possible states and adjacency matrix $[U]_{N\times N}$, which is the amount of uncertainty one can observe in accessing the states of the network \cite{Zenil}, can be measured by, $\displaystyle\mathcal{H}[U(G)]=-\sum_{i=1}^{N}p[U(x_i)]lnp[U(x_i)]$ \cite{Shannon}. The maximum value of $\mathcal{H}$ of $U$, which have $N^2$ elements $\{x_i;i=1,2,...,N^2\}$ and $x_i=0, 1 (binary)$, is given by, $\mathcal{H}_{max}=log_2[N^2]$ from which one can estimate the information contain in the network $G$ by, $I=\mathcal{H}_{max}-\mathcal{H}$ \cite{Kolmogorov}. There are various ways to calculate entropy of a network, namely, graph entropy \cite{Lian} : define probability of a node $n$ to have inner links $e$ as $p_i(n)=\frac{e}{|N(n)|}$ and probability of this node $n$ to have outer links by $p_o(n)=1-p_i(n)$, then $\mathcal{H}(G)=\sum_{n\in V}h(n)$, $h(n)=-p_i(n)lnp_i(n)-p_o(n)lnp_o(n)$, such that, $I=\mathcal{H}_{max}-\mathcal{H}(G)$. Next method is to calculate signaling entropy \cite{Teschendorff} for PPI network with each node represented by gene expression data, where, the Pearson correlation value between any pair of nodes $(i,j)$ $C_{ij}$ can be related to \textit{weight} $w_{ij}$ assigned to the edge by $w_{ij}=\frac{1}{2}[1+C_{ij}]$ from which one can obtain probability distribution $\displaystyle p_{ij}=\frac{w_{ij}}{\sum_{j\in N(i)}w_{ij}}$ to calculate entropy and information contained $\displaystyle\mathcal{H}_i[G]=-\frac{1}{log (k_i)}\sum_{i=1}^{N}[p_{ij}]log[p_{ij}]$ and $I=\mathcal{H}_{max}-\mathcal{H}$. On the other hand, simple network centrality, namely, betweenness centrality (BC) metric can be used as a significantly sensitive parameter to measure the capability of a node to spread information in a complex network \cite{Newman2}. The reason could be due to the fact that nodes having high centrality values can reach the other nodes in the network on short paths \cite{Barthelemy} enhancing the speed of information processing, and vice versa. This BC can then be correlated with the connectivity (degree) $k$ in the network as a power law $C_B(k)\sim k^{\epsilon}$ \cite{Goh}. From this relation, one can estimate maximum BC as $C_B^{max}\sim k_{max}^{\frac{\epsilon}{\gamma-1}}$, where, maximum connectivity is a function of network size $N$ by, $k_{max}\sim N^{\frac{1}{\gamma-1}}$, and $\gamma$ is the exponent of degree distribution with $\gamma=[2,3]$ \cite{Barthelemy}. Hence, centrality measurement metric can be used to characterize information processing of a node in the network. However, study on the changes in the topological properties in cancer systems in terms of the variations in information processing in the whole networks and at various levels of network organization at disease state are not studied properly in detail.\\

{\noindent}Hubs in a complex network have interesting contrast roles depending on the nature of the network, for example, if the network follows scale free then hubs try to control and manipulate the whole network such that their removal cause network breakdown \cite{Jeong1}, whereas, if the network obeys hierarchical features, hubs are busy in correlating nodes in its own module/community as well as with the communities such that their removal should not cause network breakdown \cite{Ravasz,Barabasi}. These hubs are heterogeneous in nature and can be classified into various types depending on their participation in the network and roles \cite{Guimera}. Hence, for a particular disease, it is quite important to categorize the identified key/driver regulating genes/proteins to understand their roles and potential in controlling the disease. Then, the key regulators should be studied systematically in terms of their binding affinity to various available drugs and medicinal compounds for possible drug discovery.\\

{\noindent}In this work, we address structural behavior of the OC network to understand the information processing and organization of the network along with the nature of the hubs with their classification using network theoretical approach. The behavior of the constituent communities at various levels of organization in order to understand how OC network works at fundamental level. Then the changes in the gene expression levels of the key regulators of the OC and at their stages of the cancer. Then, the key regulators are allowed to interact with the FDA approved available drugs to identify drug which could be used to prevent/cure OC disease.\\

\vskip 0.5cm
\noindent\textbf{\large Results}\\
We now present our analysis of the ovarian cancer (OC) data we have mined for study. Our main focus is three fold, first, we calculated topological properties of the complex network we have constructed using experimentally verified list of genes/proteins (\textit{Methods}) to capture the changes in the calculated topological properties at cancer state and their roles in the network organization. Second, the roles of the identified novel key regulators in the cancer state, specially, in signal transduction from information theoretical approach. Third, search for novel natural compounds (drug (FDA approved) or food nutrients) with respect to the identified key regulators.\\

\noindent\textbf{Theory of fractal properties of the ovarian cancer network}\\
{\noindent}We considered experimentally verified pre-processed 4818 genes with 16,320 interactions collected from the patients' data mined from the ovarian cancer dedicated and related database (see the workflow and \textit{Methods} for detail information of mining and pre-processing the data). The complex OC network is analyzed using Grivan and Newman community finding method \cite{Girvan}, and found that the network has five levels of hierarchical organization of communities/sub-communities till triangular motif level (Fig. 2), which are the basic functional unit of network organization \cite{Milo} and quite abundant in real networks \cite{Mangan,Alon}, and follows interesting systems level hierarchical organization in the network \cite{Malik} as also supported by the topological results of the network and its communities.\\

{\noindent}Let us represent the hierarchical scale free OC network by a graph of size $N$ $G(V,E)$, where, $V=\{n_i;i=1,2,...,N,n_i\in V\}$ is the set of nodes, and $E=\{e_{ij};i,j=1,2,...,N, e_{ij}\in E\}$ is the set of edges. The topological properties of the network, namely, probability of degree distribution $P(k)$, clustering coefficient $C(k)$, connectivity $C_N(k)$, centrality measurement parameters, betweenness $C_B(k)$, closeness $C_C(k)$ and eigenvector $C_E(k)$ centralities follow power law properties,
\begin{eqnarray}
&&P(k)\sim k^{-\gamma};~~C(k)\sim k^{-\alpha};~~C_N(k)\sim k^{-\beta}\nonumber\\
&&C_B(k)\sim k^{\epsilon};~~C_C(k)\sim k^{\kappa};~~C_E(k)\sim k^{\delta}
\end{eqnarray}
where, the power law exponents have their own values. We also found that the data of each topological distribution function of the primary network and the constituent communities collapse into a single curve after interpolation of the data following the MacKinnon and Kramer's numerical procedure which indicates that each topological distribution function ($U(k)\rightarrow P/C/C_N/C_B/C_C/C_E$) from primary network to communities/sub-communities down to motif level follow one parameter scaling law \cite{MacKinnon} given by,
\begin{eqnarray}
\label{fractal}
\frac{U}{k^{\lambda}}=H\left[\frac{\eta}{k^{\lambda}}\right]\rightarrow constant
\end{eqnarray} 
Here, $\eta$ was calculated by fitting with the scaled data and found to be approximately proportional to the minimum path length of the corresponding network/community/sub-community, and we found that $H\rightarrow constant=b$. This shows that $U(k)\sim b k^{\lambda}$, such that $\mathcal{F}(k)=\frac{U}{b}\sim k^{\lambda}$. If we take $b$ as the normalization constant such that $b=\displaystyle\sum_{i}^{N_j}U(k_i)$, where, $N_j$ is the size of the jth community/sub-community with $j=0$ corresponds to primary network $N_0\rightarrow N$, then, $0\le \mathcal{F}(k)\le 1$ and $\mathcal{F}$ behaves just like probability distribution function, such that, $\lambda=\{-\gamma,-\alpha,-\beta,\epsilon,\kappa,\delta\}$ corresponding to $P(k),C(k),C_N(k),C_B(k),C_C(k),C_E(k)$ respectively. Now if $\hat{T}$ is an operator which coarse grain the system with a scale factor $\omega$, then it is clearly seen that: $\displaystyle\hat{T}\mathcal{F}(x)=\mathcal{F}(\omega x)=(\omega x)^\lambda=\omega^\lambda\mathcal{F}(x)$ \cite{Calvert}. If one consider the property $\mathcal{F}(k)$ of the primary OC network and the same scaled property $\mathcal{F}^\prime(k^\prime)$ of any one of the communities/sub-communities of the network, then one parameter scaling law \eqref{fractal} provide the renormalization relation $\mathcal{F}(k)dk=\mathcal{F}^\prime(k^\prime)dk^\prime$: $k\rightarrow k^\prime:k=\omega k^\prime$ which gives us the following renormalization relation,
\begin{eqnarray}
\label{renorm}
\mathcal{F}^\prime(k)=\omega^{\lambda-1}\mathcal{F}(k)
\end{eqnarray}
This means that the topological properties of the OC network remains the same if the network is scale up or down by a scale factor $\omega$ which carries the fractal properties in the network organization \cite{Calvert} indicating self-organizing behavior in the OC network organization \cite{Kauffman1}. \\

\noindent\textbf{Theorem 1:} \textit{The communities which organize complex hierarchical OC network exhibit self-similar or fractal topological properties.}\\
\textbf{Proof:} Since the OC network $G(V,N)$ is hierarchical scale free topology, the network can be decomposed into $M$ communities $g_{i}(V_i,E_i); i=1,2,...,M;\forall V_i\in V,\forall E_i\in E$ with random size and structures (one can use various community finding methods, namely, Girvan and Newman method \cite{Girvan}, Louvain method \cite{Blondel}, constant potts model method \cite{Traag} and other related methods \cite{Fortunato}). The topological properties of these communities $\mathcal{F}_i$ follow one parameter scaling behavior \eqref{fractal} with the primary OC network and obeys renormalized scaling law \eqref{renorm}. Generally, these communities can be dependent on different scaling parameters $\omega_i$, such that, $g_i(V_i,E_i)\rightarrow \mathcal{F}_i=\mathcal{F}_i(k_i)$ with $k_i=\omega_ik$. Then, one can write: $\mathcal{F}_i(\omega_ik)\sim k^\lambda\mathcal{F}_i(\omega_i)$ with the scaling $k\rightarrow k_i:k=\omega_i k$. Now using renormalization relation, $\mathcal{F}(k)dk=\mathcal{F}_i(\omega_ik)dk_i$, we have the following scaling law,
\begin{eqnarray}
\label{fc}
&&\frac{\mathcal{F}_i(\omega_ik)}{\mathcal{F}(k)}=\omega_i^{1-\lambda}\nonumber\\
&&\frac{\mathcal{F}_i(\omega_ik)}{\mathcal{F}_j(\omega_jk)}=\frac{\mathcal{F}_i(\omega_i)}{\mathcal{F}_j(\omega_j)}=\left[\frac{\omega_i}{\omega_j}\right]^{1-\lambda}
\end{eqnarray}
This scaling law indicates that the topological properties of the communities of OC network have fractal properties with the primary network as well as among the communities also following the scaling law in $'k'$ given by equation \eqref{fc}. The scaling law still holds true for the organization of smaller communities at various levels of organization till motif level.\\

{\noindent}Numerical proof of the theorem can be done by the analysis of the topological properties OC network from primary network down to the motif level (Fig. 2). The behavior of $P(k)$ as the function of $k$ for different networks at various levels of organization (primary, communities, sub-communities down to motif level as shown in Fig. 2 b) are approximately similar as shown by the nearly parallel fitted lines to the data points, where, the calculated values of $\gamma$s for all networks are approximately the same (Fig. 2 a.). This clearly indicates that all the $P(k)$ data are collapsed to a single curve obeying the one parameter scaling behavior given by equation \eqref{fractal}. Similar one parameter scaling nature can be found in the behaviors of $C(k)$ and $C_N(k)$ (Fig. 2 a middle and lower panels). \\

{\noindent}The one parameter scaling property can also be observed in the behaviors of centrality measurement parameters ($C_B(k),~C_C(k)$ and $C_E(k)$) where we discussed the behaviors of high degree nodes in the respective networks (Fig. 2 c all panels). The reason is that signal processing in each respective network is mainly regulated by high degree hubs \cite{Albert,Barabasi}. However, there is significant roles of the low degree nodes in regulating the network behavior which we will be discussing also in this work. Then the data of the constituent networks fall in a single curve obeying one parameter scaling law given by \eqref{fractal}.\\

{\noindent}Lastly, we calculated network $connectance$ or $density$ \cite{Newman1} given by, $\Omega=\frac{2E_N}{N(N-1)}\sim\frac{\bar{k}}{N}$, where, $E_N$ is the total number of edges in the network of size $N$ and $\bar{k}$ is the mean degree of the network. If the network is dense then $\displaystyle\lim_{N\rightarrow\infty}\Omega\rightarrow constant$, otherwise, the network is sparse, where, $\displaystyle\lim_{N\rightarrow\infty}\Omega\rightarrow 0$. We calculated $\Omega$ for all communities at various levels of organization (Fig. 2 b) and found that all the data points collapsed to a single curve which obey power law behavior, $\Omega(N)\sim N^{-1}$ with good goodness of fit. The behavior can be seen in two ways, first, $\Omega$ rapidly decreases as $N$ increases till $N\sim 100$ then becomes nearly steady state which approximately converges to small value, but does not go to zero. This indicates that the complex OC network is weakly compact as $N$ increases. Second, since the data points beautifully falls in a single curve which obey power law behavior, the wiring/rewiring of the edges among the nodes of the communities/sub-communities is done proportionately to keep the self-similarity in the topological properties of the network.\\

\noindent\textbf{Theorem 2:}\textit{Hierarchical scale free OC network does not obey centrality lethality rule.}\\
\noindent\textbf{Proof:} Communities at a certain level of organization are loosely connected to minimize wiring energy cost to maximize information processing among them \cite{Gallos}, but within each community nodes are connected densely with large within-community clustering coefficients and significantly small path lengths \cite{Bassett} to allow each individual community function independently \cite{Sporns}. Hence, we assume that the topological properties of the communities are nearly independent of each other and we consider the network $G(V,E)$ has $M$ communities at a certain level of organization, such that, 
\begin{eqnarray}
\label{graph}
G(V,N)=\bigcup_{i=1}^{M}g_i(V_i,E_i),~~~g_i(V_i,E_i)\cap g_j(V_j,E_j)=\phi ~~~(disjoint ~sets)~~~\forall i,j\in G.
\end{eqnarray}
Now, if we consider degree $k$ as random variable which is integrable in the set $G(V,N)$, and since the the topological distribution function $\mathcal{F}(k)$ has the property $\displaystyle\sum_{i=1}^{M}\mathcal{F}_i(k_i)=1$, one can assume that this function behaves like probability distribution function. Taking distribution function of equation \eqref{graph} and applying theorem of addition of probability \cite{Gardiner}, we have,
\begin{eqnarray}
\label{prob}
&&\mathcal{F}[G(V,E)]=\mathcal{F}\left[\bigcup_{i=1}^{M}g_i(V_i,E_i)\right]=\sum_{i=1}^{M}\mathcal{F}\left[g_i(V_i,E_i)\right]\nonumber\\
&&\mathcal{F}(k)=\sum_{i=1}^{M}\mathcal{F}_i(k_i)
\end{eqnarray}
Now, if most important hub/hubs in the OC network is is/are removed, the most probable changes the perturbation can make in the network is that the community/communities $g_{c+1},g_{c+2},...,g_{c+n}$ in which the nodes present it/them will be affected the most causing rewiring of the nodes in the network \cite{Fortunato}, and in the worst scenario the community or some of the communities may collapse loosing all corresponding topological properties \cite{Heylighen}. Hence, even though the affected communities become breakdown due to the perturbation, then, corresponding topological properties will be vanished: $\sum_{j=c+1}^{c+n}\mathcal{F}(k_j)\rightarrow 0$, the other remaining $(M-n)$ communities will be survived because of their tight interaction keeping themselves \textit{small worldness property} \cite{Bassett,Sporns}, and they will reorganize the OC network with the scaling transformation of the random variable $k$ by $k\rightarrow k^\prime:k^\prime=\theta k$ and $k_i$ by $k_i\rightarrow k_i^\prime:k_i^\prime=s_ik_i=s_i\omega_ik$ so that $\mathcal{F}_i(k_i^\prime)=\mathcal{F}_i(s_i\omega_ik_i)=(s_i\omega_i)^{1-\lambda}\mathcal{F}_i(k)$ as given below,
\begin{eqnarray}
\label{com}
\mathcal{F}(k^\prime)&=&\mathcal{F}(\theta k)=\theta^\lambda\mathcal{F}(k)\noindent\\
&=&\sum_{j=1}^{M-n}\mathcal{F}_j(k_j^\prime)=\sum_{j=1}^{M-n}(s_j\omega_j)^{1-\lambda}\mathcal{F}_j(k)
\end{eqnarray}
The above equation indicates that removing hub/hubs below a critical number of hubs does not cause the collapse of the OC network even though few communities are broken down due to the perturbation, but the network reorganizes with modified topological properties. Hence, emergence of tightly bound nearly independent communities/sub-communities are the main reason in keeping complex network integration, stability, adaptation and management of external and internal perturbations and most importantly self-organization of the network, where, the activities of the leading hubs are well regulated by the emergent functional communities/sub-communities.\\

{\noindent}Proof from the OC network data analysis is as follows. We carry out hubs knock-out experiment by systematically removing the first 10-400 leading hubs from the primary OC network, then studied the changes in the topological properties of the resulting networks and compared with those of primary OC network (Fig. 3 and Fig. 4). We found the following observations. The systematic removal of the hubs from the primary OC network upto certain critical hub number $N_C^T$, the fractal properties of the resulting networks strictly kept preserved with change in the power of the topological distribution function: $\mathcal{F}(k)\rightarrow\mathcal{F}^\prime(k^\prime):$ $k\rightarrow k^\prime;~\lambda\rightarrow\lambda^\prime$. However, if the number of hubs removed is larger than this critical value, $N_C\rangle N_C^T$ then the $k^\prime$ of the resulting networks become random such that the $\mathcal{F}$ become randomly varied. For example, the exponent of power law behavior of $P(k)$ become $\gamma\rangle 3$ and $\gamma$ show random variation (Fig. 3 left panel and inset figure). For the case $N_C\langle N_C^T$ and depending on the values $\gamma$ one can classify as : (a) \textit{hierarchical nature} for $\gamma\langle 2ln2$; (b) \textit{scale free nature} for the regime $2ln2\langle\gamma\langle 3$, and (c) random for $\gamma\rangle 3$. For clustering coefficient after $N_C\rangle 80$ it is difficult to see the behavior because the calculated clustering coefficient values for those resulting networks from which large number of hubs are generally small and nearly overlapped and fitting to those data becomes quite meaningless (Fig. 3 middle panel). The scenario is distinctly visible in the behavior of neighborhood connectivity, where, the power law exponent $\beta\rightarrow 0$ for $N_C\rangle N_C^T$ otherwise finite even though there is quite fluctuations in the values (Fig. 3 panel 4 and inset). In the case of centrality measurement functions also the exponent of the power law distributions are either goes to zero ($\epsilon,\kappa\rightarrow 0$) or become random (for $\delta$) for $N_C\rangle N_C^T$ (Fig. 4 and insets). The goodness of the fit to each data measured by $r^2$ value is also quite reasonably good ($0.97\rangle r^2\rangle 0.7$). From these results we can conclude the following behavior of $\mathcal{F}$,
\begin{eqnarray}
\label{power}
\frac{\mathcal{F}^\prime(k^\prime)}{\mathcal{F}(k)}&=&\frac{k^{\prime\lambda^\prime}}{k^\lambda}~~~~~~~for~N_C\langle N_C^T\nonumber\\
&\rightarrow&0~~~~~~~~~~~for~N_C\rangle N_C^T
\end{eqnarray}
The scale free network which is obtained here is categorized based on the value of $\gamma$ ($2\langle\gamma 3$), however, the rewired network still have strongly modular structure in the network, where, $C(k)$ and $C_N(k)$ still exhibit power law behaviors. This modular structure becomes vanishing as $\gamma\rightarrow 3$, and then as $\gamma\rangle 3$ the topological properties become randomly distributed.\\

\noindent\textbf{Proposition:} \textit{Self-similarity process in OC network is given by the following scaling law:}
\begin{eqnarray}
\frac{\mathcal{F}_c(k_c)}{\mathcal{F}(k)}&\sim&\langle k_c\rangle^{-D_k};\\
\lambda&=&1+\rho\left[\frac{d_c}{d_k}\right]
\end{eqnarray}
\textit{where, $d_c$ and $d_k$ are network diameters in $c$th community, and $\rho$ is proportionality constant.}\\
\noindent\textbf{Proof:} The average diameter, which is the shortest distance between two nodes in a network, can be taken as network metric and has the form in hierarchical network \cite{Jung,Chung,Boccaletti} : $\displaystyle d\sim\frac{ln(N)}{ln(\langle k\rangle)}$. Then, one can directly write as $N\sim\langle k\rangle^{d}$, where, $N$ is the network size from which scaling law of $c$th community of the network is given by,
\begin{eqnarray}
\label{dim}
N_c\sim\langle k_c\rangle^{D_c};~~D_c\sim f(d_c)\sim f\left(\frac{l_c}{l_0}\right)
\end{eqnarray}
where, $D_c$ is the fractal dimension of $c$th community which can be related to the diameter of the $c$th community of the network. $l_c$ and $l_0$ are shortest length between two nodes in $c$th community and characteristics length respectively. Here, $\langle k_c\rangle$ is taken as the scaling parameter to understand scaling law in complex OC network. Now network mass of the $c$th community can be obtained as \cite{Song},
\begin{eqnarray}
\label{mc}
\langle M_c\rangle=\frac{N}{N_c}\sim\langle k_c\rangle^{D_c}
\end{eqnarray}
where, scaling of $k$ to $k_c$ with scaling factor $\omega[\langle k_c\rangle]$ is given by renormalization theory given below \cite{Song}, $k\rightarrow k_c:~~k_c=\omega[\langle k_c\rangle]k;~~\omega[\langle k_c\rangle]\sim\langle k_c\rangle^{-D_k}$. Renormalization theory allows to express the distribution function $\mathcal{F}(k)$ in terms of the number of nodes having 'k' links $n(k)$ in primary network with the number of nodes having $k_c$ in $c$th community $n_c(k_c)$ as, $n(k)dk=N\mathcal{F}(k)=n_c(k_c)dk_c=N_c\mathcal{F}(k_c)dk_c$. Putting the power law distribution of $\mathcal{F}(k)\sim k^{-\lambda}$ and scaling law in $\omega$ and equation \eqref{mc} we have,
\begin{eqnarray}
\label{reqn}
\frac{N}{N_c}\sim\omega^{1-\lambda}\sim\langle k_c\rangle^{D_c}
\end{eqnarray}
The fractal dimensions $D_c$ and $D_k$ given in equation \eqref{dim} can be expressed as: $D_c\sim f(d_c)\sim \omega_c d_c$ and $D_k\sim f(d_k)\sim \omega_k d_k$. Now, by using scaling in $\omega$ and equation \eqref{reqn}, one can reach the following relation,
\begin{eqnarray}
\label{lamb}
\lambda=1+\rho\left[\frac{d_c}{d_k}\right];~~\rho=\frac{\omega_c}{\omega_k}
\end{eqnarray}
The equations \eqref{dim} and \eqref{lamb}, we could relate the fractal dimension to the diameter of the network and mean degree $\langle k\rangle$, which is dependent on various network topological properties, can be taken as a scaling parameter to understand fundamental laws of network properties and their organization. Further, the distribution function $\mathcal{F}$ can be obtained as follows,
\begin{eqnarray}
\frac{\mathcal{F}_c(k_c)}{\mathcal{F}(k)}\sim\frac{k_c^{-\lambda}}{k^{-\lambda}}=\omega[\langle k_c\rangle]=\langle k_c\rangle^{-D_k}
\end{eqnarray}
The above equation indicates that if one knows the mean value of degree of a network and network diameter one can able to know the fractal properties of a network.
\\

\noindent\textbf{Significant role of lower degree hubs at cancer state in OC network}\\
{\noindent}One way to understand the role of hubs in regulating a complex network and its constituent communities/sub-communities is to study the network behavior as a function of degree ($k$) because $k$ relates directly with various important topological distribution functions \cite{Barabasi,Albert,Newman1}. One of the most important functions of the hubs involved in a complex network is the capability of triggering information flow among the nodes directly or importantly influencing the information flow among the nodes which are not connected directly, and can be characterized by behavior of centrality measures \cite{Freeman}. The behavior of betweenness centrality $C_B$ as a function of $k$ is two-fold in OC network (Fig. 4 left panel). First, the primary network and the networks in which significant number of hubs are not removed (number of highest degree hubs removed$\langle$150) $C_B$ follows two behaviors, (a) low degree nodes follow power law with larger exponent $\epsilon$: $C_B^{[s]}(k)\sim k^{\epsilon_s}$, whereas, high degree hubs obey power law with smaller value of $\epsilon$ $C_B^{[h]}(k)\sim k^{\epsilon_h}$, such that, $\epsilon_h\rangle\epsilon_s$ (Fig. 4 left panel inset). This indicates that lower degree nodes can able to communicate others in efficient manner with shorter paths which may be due to the possibility of these small degree nodes may be segregated with smaller paths \cite{Clark,Newman2}. On the other hand, high degree nodes have longer paths as compared to small degree nodes, which could be due to sparsely distribution of high degree nodes buried inside the network and proper link needed to have by each of them might be reduced significantly at cancer state. Now, one can have two sets of nodes of the whole network, $V_s=\{n_i:i\in N_s\}$ for small degree nodes, and $V_h=\{n_i:i\in N_h\}$ for high degree nodes such that $N_s+N_h=N$ and we can assume $V_s\cap V_h\sim\phi$, so that, $V=V_s\cup V_h$. It has been shown that for large complex distribution of betweenness centrality follows power law: $P(C_B)\sim C_B^{-\theta}$ \cite{Barthelemy}, hence, the joint probability distribution of betweenness centrality for the whole network can be written as, $P(V_s\cap V_h)=P(V_s)P(V_h)$ and one can have $P(V_s\cap V_h)\sim P(C_B)$. Now, we have,
\begin{eqnarray}
\label{theta}
P(C_B)&=&\int dkP(k)\delta(C_B-k^{\epsilon_s})\times \int dkP(k)\delta(C_B-k^{\epsilon_h})\nonumber\\
&\sim&\int dkk^\gamma\delta(C_B-k^{\epsilon_s})\times \int dkk^{\gamma}\delta(C_B-k^{\epsilon_h})\nonumber\\
&\sim&C_B^{-[2+\left(\frac{1}{\epsilon_s}+\frac{1}{\epsilon_h}\right)(\gamma-1)]}\nonumber\\
\theta&\sim&\left[1+\frac{1}{2}\left(\frac{1}{\epsilon_s}+\frac{1}{\epsilon_h}\right)(\gamma-1)\right]
\end{eqnarray} 
This equation \eqref{theta} shows the connection among the power law exponents $\theta$, $\epsilon_s$, $\epsilon_h$ and $\gamma$. Now, as the number of high degree hubs are removed $\epsilon_s\rightarrow\epsilon_h$, and at a certain critical number of hubs removed $N_c^{[t]}~(\sim 200)$ where the two fitted power law in $C_B$ merges $\epsilon_s=\epsilon_h$ one can retrieve $\displaystyle\epsilon_s=\epsilon_h=\epsilon=\frac{\gamma-1}{\theta-1}$ \cite{Barthelemy,Vazquez}. Further, for $N_c\rangle N_c^{[t]}$, we found that $\epsilon$ decreases such that $\displaystyle\lim_{N_c\rightarrow large}\epsilon\rightarrow 0$ which indicates that the path length becomes constant indicating information flow becomes constant.\\

{\noindent}Similar behavior is found in the behavior of closeness centrality (CC) which is the measure of distance between a node to all other nodes in the network segregated with close shortest paths from it \cite{Freeman1}. The nature of closeness centrality of high degree nodes as a function of degree exhibits power law $C_C^{[h]}(k)\sim k^{\delta_h}$, but the exponent $\kappa_h$ has comparatively smaller value as compared to that of lower degree nodes, $C_C^{[s]}(k)\sim k^{\delta_s}$, $\delta_h\rangle\delta_s$ (Fig. 4 right panel). Since CC indicates the measure of geographical distance to spread the information from a node in the network \cite{Newman2,Freeman1}, the rate of information transfer coordinated by high degree hubs is significantly reduced as compared to those of low degree nodes at OC state. One of the key reasons could be the direct drastic changes in the expression of some of the driver genes (high degree hubs) in the genetic network at cancer state, where, rest of the existing nodes (mainly low degree nodes) in the network are indirectly influenced \cite{Malik,Martinez}. However, as the low degree nodes generally tie with high degree hubs \cite{Freeman2}, the low degree nodes might know the changes in the information processing of the high degree hubs in the network and probably might act accordingly to keep efficiency of information processing in the network. Further, as the number of high degree hubs removed is increased $\delta_s\rightarrow\delta_h=\delta$ and also $\displaystyle\lim_{N_c\rightarrow large}\delta\rightarrow 0$ indicating information spread becomes constant.\\

{\noindent}The nature of eigenvector centrality (EC) is also quite similar to that of betweenness centrality and obeys power law behavior, where, the exponent of the power law function of the low degree nodes of the OC network, $C_E^{[s]}(k)\sim k^{\kappa_s}$, is larger than that of high degree nodes, $C_E^{[h]}(k)\sim k^{\kappa_h}$, i.e. $\kappa_h\rangle\kappa_s$ (Fig. 4 middle panel and inset). The importance of EC is that this network metric not only reveals the influence of a node in the neighboring nodes but also the score of the connected nodes \cite{Bonacich}. Hence, the significant decrease in the value of $\kappa_h$ of the high degree hubs indicates that the signal propagation and influence of the hubs at various layers of neighbors of nodes starting from each one of them connecting till the end of the network \cite{Spizziri} is quite reduced at cancer state. Whereas, the low degree nodes, which received change in influence of the OC network, maintain faster rate where $\kappa_s$ is large as compared to $\kappa_s$. As seen before in other centrality measures we have: $\kappa_s\rightarrow\kappa_h=\kappa$ as $\displaystyle\lim_{N_c\rightarrow large}\kappa\rightarrow 0$, such that, $C_E(k)\rightarrow constant.$ at large $N_c$ limit.\\

\noindent\textbf{Provincial hubs are prominent regulators in OC network}\\
{\noindent}The behavior of neighborhood connectivity $C_N(k)$ (Fig. 3 right panel) obeyed power law nature with negative exponent which is the signature of disassortativity nature \cite{Pastor,Newman3}. This disassortativity nature, which is again supported by the rich-club coefficient $\phi$ and normalized rich-club coefficient $\phi_{N}$ results (Fig. 5), reveals that there is no evidence of rich-club formation of the high degree hubs \cite{Colizza}. Since the normalized rich-club coefficient $\phi_{N}\langle 1$ for the primary network, the high degree hubs are sparsely distributed in the OC network topology
and this value of $\phi_{N}$ is still found to be smaller for those networks after few hubs are knockout $N_c\rangle 10$ indicating the correlation among the high degree hubs become decreased significantly as more number of highest rank hubs are removed from the network \cite{van}. This weakening in the correlation among the rich degree hubs in the complex OC network reduces the number of \textit{attractors} of the nodes in the network and loosen the connectivity among the low degree and peripheral nodes in the network causing drastic functional diversity \cite{Senden}. Hence, the dominance of controlling of the network by these rich-degree hubs such signal processing, functional structuring and many other regulating mechanisms will be less probable in OC network. Even though the hubs having $k\ge 30$ have the tendency to tie tightly among themselves (increasing features of $\phi$ and $\phi_{N}$ in Fig. 5), their controlling dominance is weak ($\phi_{N}$). Since the OC network is hierarchical topology, the loosely bound low degree nodes and moderate degree hubs to the high degree hubs might have segregated at cancer state and organized forming diverse loosely connected communities/subcommunities \cite{Colizza} to maintain optimal efficiency of the system's properties. Hence, the signaling behavior of high degree hubs could be different from that of lower degree nodes ($\epsilon_s\rangle\epsilon_h$, $\kappa_s\rangle\kappa_h$ and $\delta_s\rangle\delta_h$ in Fig. 4) in order to keep network organization optimal even at cancer state.\\

{\noindent}Now, in order to identify the type of the hubs which take prominent responsibility for OC network regulation we calculated participation ratio $P_i$ of the nodes with respect to $Z_i$ score (see \textit{Methods}) showing the different types of hubs for different number of leading hubs removed (Fig. 6). The analysis of the evolution of the different types of hubs in $(P_i,Z_i)$-plane as a function of $N_c$ show that provincial hubs are more prominent as compared to other types of hubs, namely, kinless and connector hubs. For this we defined the probability of finding each type of hubs $P_H$ as the ratio of the number of each type of hub $N_u;~\forall u\in\{k,q,p\}$ ($N_k\rightarrow number~of~kinless~hubs$, $N_q\rightarrow number~of~connector~hubs$ and $N_p\rightarrow number~of~provincial~hubs$) to the size of the network $N$ as given below,
\begin{eqnarray}
P_H(N_c)=\frac{N_u}{N}\rightarrow\begin{cases}\frac{N_k}{N},&kinless~hubs\\
\frac{N_q}{N},&connector~hubs\\
\frac{N_p}{N},&provincial~hubs\end{cases}
\end{eqnarray}
From the Fig. 7 we could see that probability of finding kinless hubs is quite low as compared to other types of the hubs (connector and provincial hubs) and could able to see when one removed $N_c=25,30$ and surprisingly there is no signature of it for $N_c\rangle 30$ and $N_c\langle 25$. However, probability of finding of connector and provincial hubs could be seen starting from primary network and in the regime $0\le N_c\le 50$ connector hubs abundance dominate that of provincial hubs in the network. This indicates that even though first leading 50 hubs from the OC network are removed or their regulatory functions are suppressed due to signal manipulation at cancer state, these connector and provincial hubs take over main responsibilities to keep network integrity, stability, regulation and preserve network topological properties. As the OC severity increases (cancer state moves to stage I, II, III etc.) the number of genes expressed/suppressed start increasing \cite{Schaner} most probably causing in increase in the number of manipulated high degree hubs in the OC network \cite{Bast}. In such situation cancer driven hub manipulation, both connector and provincial hubs take major responsibility in network regulation of silencing of number of hubs $N_c\le 100$. If the number of manipulated hubs is $N_c\rangle 100$ only provincial hubs take main responsibility of network integrity and regulation.\\

{\noindent}The key regulators of OC network \cite{Malik}: CD44, EPCAM, KRAS and MCAM are found to be \textit{provincial} hubs, and AKT1 falls in \textit{connector} hub (Fig. 5b lowest panel). All the key regulators do no change their hub identity in the knockout experiment of $N_C=10$, except CD44, where, it changes from provincial to connector as around $N_C=40$ hubs are knockout from the OC network. This indicates that provincial and connector hubs are quite important in understanding OC network regulation in maintaining network stability, properties and self-organization when the network is perturbed externally and internally by cancer signal.\\

\noindent\textbf{Patterns of mRNA expressions of the key regulators show diverse activities}\\
{\noindent}Now, we studied the activities of the key genes \cite{Malik} from the mRNA expressions of the five key regulators using GEPIA (Gene Expression Profiling Interactive Analysis) online server (see \textit{Methods}) where p-value $P\rangle 0.01$. The mRNA expression levels of these key regulators show diverse activities at OC state as compared to that of normal expressions (Fig. 6a). We found that EPCAM was over expressed at ovarian cancer state, whereas, MCAM was down regulated but CD44 was significantly up regulated as compared to that of normal state. However, AKT1 and KRAS were slightly up regulated as compared to that of normal state. This is quite strongly correlated to the observed survival analysis of these key regulators \cite{Malik}, where, some regulators (AKT1, CD44, MCAM) act as cancer suppressor, whereas, some of them (EPCAM and KRAS) work in the cancer progression. Hence, expressions AKT1, CD44, EPCAM and KRAS can be correlated to increasing or decreasing the risk of disease progression, and hence, they can be potential prognostics markers or drug targets in OC carcinogenesis. Then, the mRNA expression levels of the key regulators are calculated and clearly show that at stage-II all the key genes unregulated maximally, and then slowly down regulated as the cancer stage approaches from $II\rightarrow IV$ (Fig. 6b). Moreover, another important aspect of these results is that the overall mRNA expression level EPCAM at any OC state and it's expression level at any stage of the OC is maximum, whereas, those of KRAS are minimum. Further, the expression levels of EPCAM at any stages (II-IV) of OC remain equally over expressed. Across all stages (I–III), EPCAM was the most frequently expressed biomarker \cite{Patriarca}. EPCAM could therefore be an ideal tumor antigen candidate for detecting cancer cells. Overexpression of EPCAM was connected to advanced stage of disease and poor overall survival in some tumor types, suggesting EPCAM as a possible prognostic marker. AKT activation is prevalent in high-grade, late-stage serous ovarian carcinomas (6–9), according to immunohistochemical investigations, and may thus play a role in promoting tumor development \cite{Yuan,Altomare, Cristiano,Kurose,Phung}. \\

{\noindent}The TIMER web tool \cite{Taiwen} analysis showed that the expression of AKT1, CD44, EPCAM, KRAS and MCAM genes was significantly correlated with one or more tumor-infiltrating immune subsets (Fig. 7). We found that the immune infiltrates, namely, CD4+ T-cells show positive correlations with the expressions of AKT1, CD44, MCAM and KRAS; macrophages have positive correlation with KRAS and MCAM; and neutrophils with the expression of KRAS. This indicates that the key regulators have the most significant correlation with immune infiltrates. Important aspects of KRAS is that it's expression has positive correlation with maximal infiltrates (Purity, CD4+ T cell, macrophage, neutrophil, dendritic cell). Hence, it is quite important to notice that these key regulators involve in immune infiltrates and regulations connecting to OC with immune system \cite{Yigit}.\\

\noindent\textbf{Network based drug repurposing and docking analyses}\\
{\noindent}The identified novel key regulators of OC network using network theoretical approach are shown to be potential drug targets to prevent/cure this disease because of their high connectivity with the rest of the nodes (both high and low degree) in the network and the ability to rapidly transfer information in the network. Key regulator genes could be the most appropriate target in complex biological network for drug detection because of their high connection and capacity to rapidly transfer information\cite{Prasad}. Hence, we use these five regulators for possible identification of available list of the FDA approved drugs. Now a total of 219 drug-gene interactions for these key genes were identified (Supplementary Table). We found that out of these available list of drugs, \textit{Uprosertib, Progesterone, Solitomab} and \textit{Regorafenib} drugs showed significant strong interactions with the key genes AKT1, CD44, EPCAM and KRAS respectively having the highest score (Fig. 8A). Many of the drugs targeting these key regulators can be used either individually or in combination. Uprosertib is a reversible pan-AKT inhibitor that interacts with adenosine triphosphate \cite{Pachl,Dumble,Tolcher}. A recent study demonstrated that ovulation, the release of hormones (e.g. progesterone and estrogen), inflammatory factors and growth factors among others, promoted the migration of intrauterine-injected malignant cells towards the ovarian stromal compartment to form ovarian tumors \cite{Yang}. Solitomab is an EPCAM antibody that is being developed for the treatment of patients with multiple EPCAM-expressing solid tumors \cite{Brischwein,English}. Because of its capacity to engage resting, polyclonal, CD8+, and CD4+ T lymphocytes for very potent, redirected destruction of target tumor cells, solitomab has exhibited exceptional anticancer activity in preclinical ovarian tumor xenograft models \cite{English}. Regorafenib is an anti-angiogenic oral multikinase inhibitor that has been approved for use in a variety of solid cancers \cite{Poklepovic}.\\

{\noindent}We then did molecular docking analysis of the filtered drug compounds with the AKT1, CD44, EPCAM, KRAS and MCAM proteins as shown in Fig. 8B. Based on their binding affinities and visual inspection of docked complexes for their ability to shape H-bonds and interactions key genes and ligands (Fig. 8B). Binding affinities of AKT1, CD44, EPCAM and KRAS with drug Uprosertib (-7.723 kcal/mol), Progesterone (-3.715 kcal/mol), Solitomab (-6.877 kcal/mol) and Regorafenib (-3.934) respectively, showed greatest affinity of binding among these. The key protein (AKT1, CD44, EPCAM and KRAS) is bind with drug with hydrogen bond and other non-polar bond with residues of key proteins (Fig. 3B). 

\vskip 0.3cm
\noindent\textbf{\large Conclusion and discussion}\\
{\noindent}Identification of driver genes in OC and understanding their roles are still open questions and critical study in this field is necessary to prevent this disease and also finding a possible drug target/s to cure this disease. The study can be done from diverse directions of experimental techniques, theoretical methods, data analysis etc with interdisciplinary perspective. We mined experimentally verified genes, which involved in OC, from various dedicated databases and analyzed that data from network theoretical approach to understand how OC network is organized, what is the role of the hubs, what could be the mechanism of signal manipulation in OC etc.\\

{\noindent}The OC network carries fractal properties which is reflected in the topological properties of the network as power law distributions. If one goes deeper inside the network through communities/subcommunities till basic building block motif level (OC network has five hierarchical levels of organization) one can observe self-similar power law distributions which could be the origin of fractal nature in the OC network. These communities/subcommunities can be think of as coarse grained functional modules which can be scale down to basic functional units whose functional properties are self-similar because the topological properties of these functional modules at all levels of organization obey one parameter scaling law where all the curves collapse to a single curve by interpolating the curves. We proved it from both from data analysis and general fractal theory approach \cite{Mandelbrot} and breakdown of any of the functional module caused by removing by knocking out few leading hubs from the network do not cause OC network breakdown. This property reveals that OC network does not follow centrality lethality rule \cite{Jeong1} and self-organized. \\

{\noindent}Understanding the role/s of the hubs are important because driver genes are generally one of these hubs and they are meant for potential drug target to enable to eradicate OC disease. The hubs knockout experiment we performed clearly shows us few important features in the OC network. First, removal of significant number of hubs from the OC network allows the transition from hierarchical network to scale free topology. Continued removal of hubs bring changes from the scale free topology to random feature which is the signature of network breakdown. This indicates the significant role of high degree hubs in shaping the OC network topology as well as keeping network stability. Second, one can observe the decrease in the rate of information processing of the leading hubs which could be due to the manipulation of activity of hub nodes as compared to the activity of lower degree hubs. This manipulation of activities of high degree hubs at cancer state leads to decrease in the number of 'attractors' in the network structure causing segregation of lower degree hubs with short paths to become more active. However, the nodes in the OC network try to keep the network properties, stability and organization after rewiring among them.\\

{\noindent}There are different types of hubs in OC network but surprisingly it is not the high rank hubs but moderate rank 'provincial' and 'connector' hubs which are the keepers and safeguarders of the OC network at cancer state when the cancer signal spreads all over the entire network. There is no signature of formation of 'rich-clubs' by the high rank hubs in the OC network causing weakening the ties of the lower degree hubs in the network periphery with the hubs affecting in the functional diversity of the network. Hence, there is quite complex regulatory mechanisms in network organization when the system is at cancer state. Further, when the OC moves to severity condition i.e. stage I, II, III etc which could be modeled by allowing the network to destabilize more high degree hubs \cite{Schaner} the network regulation becomes more complicated. Hence, we need to study further on the cancer data at various severity stages so that one can detect OC at early stage for prevention. Another important aspect is to study the proper analysis of these key genes and clinical trials for possible drug discovery to eradicate this cancer type.\\

{\noindent}We then studied the gene expression patterns of these key regulators at OC state and their behaviors at different stages of the OC. The analysis indicates that EPCAM has maximum over expression overall at various stages of the cancer state which indicates this gene behaves as cancer enhancer in the OC progression \cite{Zheng}. It is also found that CD44 is also highly expressed as compared to normal state and could also be another important target gene \cite{Gao}. Similarly, other remaining regulators are also important for analysis in terms of drug targets and other analysis. These regulators are also positively correlated with immune infiltrates such as CD4+ T cells etc indicating the correlation of OC with immune system. Although CD44 expression has been observed during the progression of ovarian carcinoma, the questions of whether high CD44 expression indicates a relatively favourable prognosis \cite{Kayastha} or aggressive tumor behaviour and a poor prognosis \cite{Kayastha} and whether CD44 has any prognostic significance \cite{ Berner, Saegusa, Cannistra} have remained unresolved. To better understand the significance of CD44 expression in epithelial ovarian cancer, researchers used immunohistochemistry to examine a large number of samples and examined the prognostic value of CD44 using multivariate and univariate analyses \cite{Sillanpaa}. The finding points CD44 playing a role in establishing treatment resistance by stimulating the PI3K/Akt signalling pathway \cite{Liu}. In preclinical models, an RNA-based bispecific CD44-EPCAM aptamer was created and tested. By combining single CD44 and EPCAM aptamers with a double-stranded RNA adaptor, it can block CD44 and EPCAM at the same time. In vitro and in vivo investigations, the bispecific CD44-EPCAM aptamer was much more effective than either single CD44 or EPCAM aptamer at suppressing cell proliferation or inducing apoptosis in ovarian cancer cells \cite{Zheng1}. The findings show that the KRAS-variant is a new genetic marker for OC risk, and that it could explain genetic risk in a number of hereditary ovarian cancer and breast  syndrome families, according to the researchers \cite{Ratner}. In an in vitro situation similar to the patient's tumour, proliferation is aided by MER-mediated ERK and AKT signalling, which enhances phosphorylated nuclear ER. When MAPK activity is increased, inhibiting both the Mutated, Estrogen Receptor, and ERK pathways is required to stop proliferation, which may explain the use of aromatase inhibitors rather than tamoxifen \cite{Kato}. When there are significant genomic co-alterations, such as KRAS or PI3K, it has been suggested that personalised combination therapy may be a key to establishing tumur control \cite{Sicklick}, especially when certain gene product pathways, such as KRAS or PI3K, are activated \cite{KatoS}. It could be a useful indicator for predicting ovarian cancer metastasis, recurrence, and drug resistance, as well as a diagnostic, therapeutic, and prognostic tool. However, it's surprising that the stronger key regulators expression is, the earlier the clinical stage is, implying that it's a protooncogene. However, in the context of prognosis, it also functions as a tumour suppressor gene. This issue will be the subject of our future research.\\

{\noindent}Further, network theoretical approach was used to identify potential drugs from the list of FDA approved drugs which have high affinity to bind with the key proteins and high score in molecular docking analysis. The identified drugs could be useful for the prevention of this disease and need to be done critical study on it both from theoretical and experimental perspectives. We propose a systematic investigation on the identified key/driver genes and need to do various clinical trials and experiments.

\vskip 0.5cm
{\noindent}\textbf{\large Methods}\\
\vskip 0.2cm
\noindent{\textbf{Flowchart of the method}\\
The methods and algorithms we have used to study OC network are summarized in the flowchart in the Fig. 1.\\

{\noindent}\textbf{Construction of PPI network and Knockout of nodes}\\
{\noindent}We chose six highly cited ovarian cancer databases, namely, COSMIC database, Gene cards database, Ovarian kaleidoscope database, Dragon database of ovarian cancer,
Curated ovarian database and OC- Gene database, and mined 4818 experimentally verified genes which are involved in OC. We then curated the identified list of the genes by using CGI code written in Perl and removed gene duplication and involved aliases in the list. From this method we could able to retrieve 660 genes out of 2000. This curated list of the genes is then cleaned manually as well as using Cytoscape 3.7.1 plugin by mapping to UniProt (January 2016), and we arrived at final list of 600 genes. Now following the principle of one gene one protein idea we constructed OC primary network using the curated 600 genes using GeneMANIA App \cite{Sara} by uploading the 600 genes in Cytoscape 3.7.1 \cite{Markiel}. The primary network so constructed is composed of 4818 with 16320 physical interactions among the involved nodes.\\

{\noindent}The knockout experiments of the high degree hubs were carried out by removing the first 10-400 leading hubs along with their associated edges from the primary OC network systematically and analyzed the topological properties of the resulting networks.\\

\noindent\textbf{Network topological properties}\\
{\noindent}Consider a complex network of size $N$ defined by a graph $G(V,E)$, where, $V=\{n_i; i=1,2,...,N;\forall n_i\in V\}$ and $E=\{e_{ij}; i,j=1,2,...,N;\forall e_{ij}\in E\}$ are sets of nodes and edges respectively. Since the PPI we constructed is undirected and unweighted network, $G(V,E)$ can be constructed by defining an adjacency matrix $[A]_{N\times N}$ with matrix elements, $A_{ij}=1$ if there is an edge between 'i'th and 'j'th nodes in the network, otherwise $A_{ij}=1$, and $A$ is symmetric matrix \cite{Newman1}. We used $Cytoscape 3.6.0$ with modules \textit{Network analyzer} and \textit{CytoNCA} \cite{Shannon,Doncheva,Tang} and \textit{Igraph} for calculating the topological properties of the OC network. We employed \textit{Louvain community detection algorithm} \cite{Blondel} for digging communities/subcommunities at various levels of organization embedded in OC network. The topological properties of a complex network which are calculated for analysis are the following: \\

{\noindent}\textbf{Degree ($k$):} This network metric is the total number of links a node has in a network. If $k_i$ is the degree of its node of the network, then it can be obtained by,
\begin{eqnarray}
\label{eq:1}
k_i=\sum_{j=1}^{N}A_{ij}
\end{eqnarray}
This network metric characterize how much a node can influence other remaining nodes in the network \cite{Newman1}. Larger the value of $k_i$ more influential the its node is and vice versa.\\

{\noindent}\textbf{Probability of degree distribution, $P(k)$:} It indicates the measure of the probability of finding nodes of degree $k$ in the network and can be calculated as follows, 
\begin{eqnarray}
\label{eq:2}
P(k)=\frac{n_k}{n}
\end{eqnarray}
where, $n_k$ is the total number of nodes with degree $k$ in the network. This network metric can able to characterize topological features of different types of networks, such as, scale free and hierarchical network follow power law distribution with $k$, whereas, small world or random network obey Poisson distribution with $k$ \cite{Ravasz,Albert,Barabasi}.\\

\noindent\textbf{Clustering coefficient (local) $C_i(k)$:} It is a network metric that characterize the ability of a node to have local internal links with its nearest neighbors, which can be measured by by the ratio of the number of linked pairs of neighbors of a node ($L_i$) to total number of possible pairs of neighbors of that node \cite{Newman1}. It is given by,
\begin{eqnarray} 
\label{eq:3}
C_i(k_i)=\frac{2L_i}{k_i(k_i-1)}
\end{eqnarray}
Higher the value of $C_i$ local nodes are densely connected. However, higher degree nodes generally have lower value of clustering coefficient \cite{Albert}. For hierarchical network $C_i$ can be conjectured by $C_i(k)\sim k^\alpha$ with $\alpha\le 1$, whereas, for scale free, random and small world network $C_i(k)\rightarrow constant$ \cite{Ravasz,Albert,Barabasi}. One can also calculate average clustering coefficient of the network by using, $C(k)=\displaystyle\frac{1}{N}\sum_{i=1}^{N}C_i(k)$.\\

\noindent\textbf{Neighborhood connectivity $C_N(k)$:} This metric characterize the average ties of a node with the nearest neighbours of degree $k$ in a network defined as follows \cite{Pastor},
\begin{eqnarray}
\label{eq:5}
C_N(k)=\sum_{s}sP(s|k)
\end{eqnarray}
where, $P(s|k)$ is conditional probability of the links of a node with $k$ connections to another node having $s$ connections. If the $C_N(k)$ exhibit power law behavior $C_N(k)\sim k^{\pm\beta}$ with $\beta\le 0.5$, then the network falls to have in hierarchical features \cite{Pastor}, where, $+\beta$ behavior is the signature of $assortativity$ property of the network, whereas, $-\beta$ indicates $disassortativity$ nature of the network. Networks with assortativity nature is the evidence of $rich-club$ formation of high degree hubs in the network \cite{Colizza}.\\

\noindent\textbf{Rich-club coefficient ${\phi(k)}$:} This metric is the measure of how well high degree hubs establish connections among themselves to control the network regulation, and can be defined as the ratio of the number of links connected among the rich-club member hubs ($U_{\rangle k}$) to the maximum possible links they ($M_{\rangle k}$) can have \cite{Colizza}, as given below,
\begin{eqnarray}
\label{eq:6}
{\phi(k)}=\frac{2U_{>k}}{M_{>k}(M_>k-1)}
\end{eqnarray}
Now the normalized rich-club coefficient can then be written as follows \cite{Colizza},
\begin{eqnarray}
\label{eq:7}
\phi_{norm}(k)=\frac{\phi(k)}{\phi_{rand}(k)}
\end{eqnarray}
where, $\phi_{rand}(k)$ is the calculated \textit{rich-club coefficient} of the generated \textit{random networks} with similar size and degree sequence. If the network have signature of rich-club then $\phi_{norm}(k)\rangle 1$, otherwise ($\phi_{norm}(k)\langle 1$), there is rich-club formation.  \\

\noindent\textbf{Betweenness centrality $C_B$:} It is the measure of the influence of a node has to propagate an information in the other nodes of the network \cite{Newman1,Borgatti,Freeman3} which can be quantified as follows,
\begin{eqnarray}
\label{eq:9}
C_b(v)=\sum_{i,j;i\ne j\ne k}\frac{d_{ij}(v)}{d_{ij}}
\end{eqnarray}
where, $d_{ij}(v)$ denotes the number of geodesic paths from node $i$ to node $j$ passing through node $v$ and $d_{ij}$ is the all possible number of geodesic paths from node $i$ to node $j$. Hence, normalized betweenness centrality $C_B$ is given by, 
\begin{eqnarray}
\label{eq:10}
C_B(v) = \frac{1}{M}C_b(v)
\end{eqnarray} 
where, $M$ denotes the number of node pairs, excluding $v$ Higher the value of $C_B$ faster the information processing throughout the network and vice versa.\\

\noindent\textbf{Closeness centrality $C_C$:} It can be regarded as the measure of how far an information can be propagated from a certain node to the rest of the other nodes in the network \cite{Bavelas,Newman1,Borgatti}. This metric of any node $i$ can be calculated from the harmonic mean of the all possible geodesic paths $d_{ij}$ linking to the node $i$,
\begin{eqnarray}
\label{eq:8}
C_C(k)=\frac{N}{\sum_{j}d_{ij}}
\end{eqnarray}
Hence, larger the value of $C_C$ farther the information can be reached, and vice versa \cite{Newman1}.\\

\noindent\textbf{Eigenvector centrality $C_E$:} This network metric is the measure of strength of influence of a node to the rest of the nodes in the network having diverse influential capabilities \cite{Bonacich}. The eigenvector centrality of a node $i$ in a network can be obtained,
\begin{eqnarray}
\label{eq:11}
C_E(i)=\frac{1}{\lambda}\sum_{j=nn(i)}{v_j}
\end{eqnarray}
where, $nn(i)$ represents the  nearest neighbors of node $i$ in the network with eigenvalue $\lambda$ and eigenvector $v_i$ in the eigenvalue equation, $Av_i={\lambda}v_i$ where, $A$ is the network adjacency matrix. $C_E$ values of all nodes in a network are always positive $C_E\rangle 0$, and it is considered to be the extension of degree centrality \cite{Newman1}.\\

\vskip 0.3cm
{\noindent}\textbf{Techniques to classify \textit{hubs}} \\
Classification of different types of \textit{hubs} in a network can be done by the following network metrics.\\

\noindent\textbf{Z-score parameter ($Z_i$):} This parameter is generally known as within module Z-score parameter which measures how strongly a particular node in a module is linked to the rest of the nodes in the module \cite{Guimera}. The Z-score of a node $i$ having $k_i$ links in certain module "$c$" can be defined as,
\begin{eqnarray}
\label{eq:16}
Z_i=\frac{k_i-\langle k\rangle_{c}}{\sigma_{k_{c}}}
\end{eqnarray}
where, $\langle k\rangle_{c}$ and $\sigma_{k_{c}}$ are the expectation value and standard deviation of degrees $k$ of all nodes in the module respectively. This $Z-$score can able to distinguish hubs from non-hub nodes depending on its value, namely, $modular-hubs$ if $Z_i\ge 2.5$ (nodes having large number of links) and $non-hubs$ if $Z_i\le 2.5$ (lower degree nodes) \cite{Guimera}.\\

\noindent\textbf{Participation co-efficient ($P_{i}$):} The local and global roles of the identified modular hubs by $Z$-score can be further classified depending on the participation of the hubs locally within their own modules and globally by establishing links to other modules other than its own module. This participation co-efficient of node $i$ in a network having $N_M$ modules, which have degree $k_i$, can be defined as \cite{Guimera},
\begin{eqnarray}
\label{eq:17}
P_i=1-\sum^{N_M}_{s=1}\left(\frac{k_{ic}}{k_i}\right)^2
\end{eqnarray}
where, $k_{is}$ denotes the degree of node $i$ in $c$ module. If $P_i\rightarrow 1$ then the node $i$ is uniformly linked to all other modules, whereas, if $P_i\rightarrow 0$ the node links to the nodes within its own module \cite{Guimera}. Generally, nodes which have $P_i\rangle 0.7$ has larger participation in other modules. Now, different types of hubs can be classified depending on the values of $Z$-score and $P_i$ in $(P_i,Z_i)$-plane as follows: 1. \textit{kinless hubs:} Hubs in the range $Z_i \ge 2.5$ \& $P_i  > 0.70$, 2. \textit{connector hubs:} in the area $Z_i \ge 2.5$ \& $0.30 < P_i \leq 0.75$, and 3. \textit{provincial hubs:} in the area $Z_i \ge 2.5$ \& $P_i \leq 0.30$ in the $(P_i,Z_i)$ space \cite{Guimera}.\\

\noindent\textbf{Validation of the expression pattern of key genes}\\
\textcolor {blue}The GEPIA \cite{Tang1} for Cancer Genomics was used to investigate the expression pattern of the key genes. Tumor or normal differential expression analysis, profiling according to cancer subtypes or pathological stages, patient survival analysis, similar gene finding, correlation analysis, and dimensionality reduction analysis are all customizable functions in GEPIA\cite{Tang1}.}\\

\vskip 0.3cm
\noindent\textbf{Correlation of gene expression with tumor-infiltrating immune cells:}\\
Considering the importance of immune dysregulation in ovarian cancer, we explored the correlation between AKT1, CD44, EPCAM, KRAS and MCAM mRNA expression and tumor-infiltrating immune cells. The web tool TIMER (https://cistrome.shinyapps.io/timer/) \cite{Taiwen} was used. Six tumor-infiltrating immune subsets, neutrophils, macrophages, CD8+ T cells, CD4+ T-cells and including B cells were analysed.\\

\vskip 0.3cm
\noindent\textbf{Network-Based Drug Repurposing:}\\
The DGIdb \cite{Cotto} database was screened out for discovering therapeutic targets for key genes. The DGIdb was used to produce a similar drug-gene interaction network. DGIdb is a database that can anticipate the physical and functional interactions between key genes and drugs in a query. The interactions in the DGIdb database come from five different sources: automatic high-throughput lab experiment data, text mining,  co-expression interaction data, genomic context interaction prediction, and previous knowledge from other databases.\\

\vskip 0.3cm
\noindent\textbf{Protein structure preparation:}\\
The X-ray structures of some key genes of ovarian cancer like AKT1, CD44, EPCAM, KRAS and MCAM are available on protein data bank (PDB) \cite{Berman}corresponding to PDB ids 9S9W, 1POZ, 4MZV, 4EPT and 6LYN, respectively. Prior to docking studies, the proteins were processed using Schrodinger Protein Preparation Wizard during which missing hydrogens were added, hydrogen bonds were optimized and water molecules were deleted \cite{ Madhavi}.\\

\vskip 0.3cm
\noindent\textbf{Ligands preparation:}\\
Nutraceuticals are bioactive phytochemicals that deliver health benefits and are relatively safe to be used for prevention and treatment against disease.  A large number of drugs are FDA approved pharmaceutical drugs. The chemical structures of high score ligands from gene-drug interaction of key genes were obtained from PubChem \cite{kimS}. The chemical structures of ligands were prepared with LigPrep module Schrodinger which generated diverse, accurate and energy minimized conformations using OPLS-2005 force field \cite{ Madhavi}.\\

\vskip 0.3cm
\noindent\textbf{Docking studies using Glide:}\\
Glide uses a range of filters and performs a thorough search of the conformational, orientational and positional space for ligand in the binding site of receptor. The receptor was represented by creating a cubic grid of size 30x30x30 Å centred on the active site residues for each protein by Schrodinger’s Receptor Grid Generation program \cite{Halgren}. Further the ligands were screened against each receptor using high throughput virtual screening (HTVS) approach of Glide \cite{Halgren}. The high ranking ligands were then subjected to stringent screening via extra precision (XP) \cite{Halgren} method to eliminate false positives and acquire precise binding modes of compounds. 
\vskip 0.3cm
\noindent\textbf{Interaction analyses}\\
Maestro was used to compute interactions between the docked protein-ligand complexes. 2D diagrams were generated depicting hydrogen bonds and hydrophobically interacting residues. Then PyMOL \cite{Seeliger} was used to visualize and study the resultant receptor-ligand complexes.\\
\vskip 1.5cm
\noindent\textbf{Acknowledgments}\\
KC and RKBS are financially supported by UPE-II, under sanction no. 101 and DBT-COE. MZM financially supported by DHR, Indian Council of Medical Research under Young Scientist Research Fellowship).
\vskip 0.5cm
\noindent{\bf Author Contributions} \\
RKBS conceived the model. KC, MZM and RKBS did the numerical experiment. KC, MZM and RKBS prepared the figures of the numerical results, analyzed and interpreted the simulation results. RKBS, KC and MZM wrote the manuscript. KC, MZM, PS and  RKBS involved in the study and reviewed the manuscript.

\vspace{0.5cm}
\noindent {\bf Additional Information} \\
\textbf{Competing financial interests:} The authors declare no competing financial interests.

\newpage
\begin{figure}
\label{fig1}
\begin{center}
\includegraphics[width=1\textwidth]{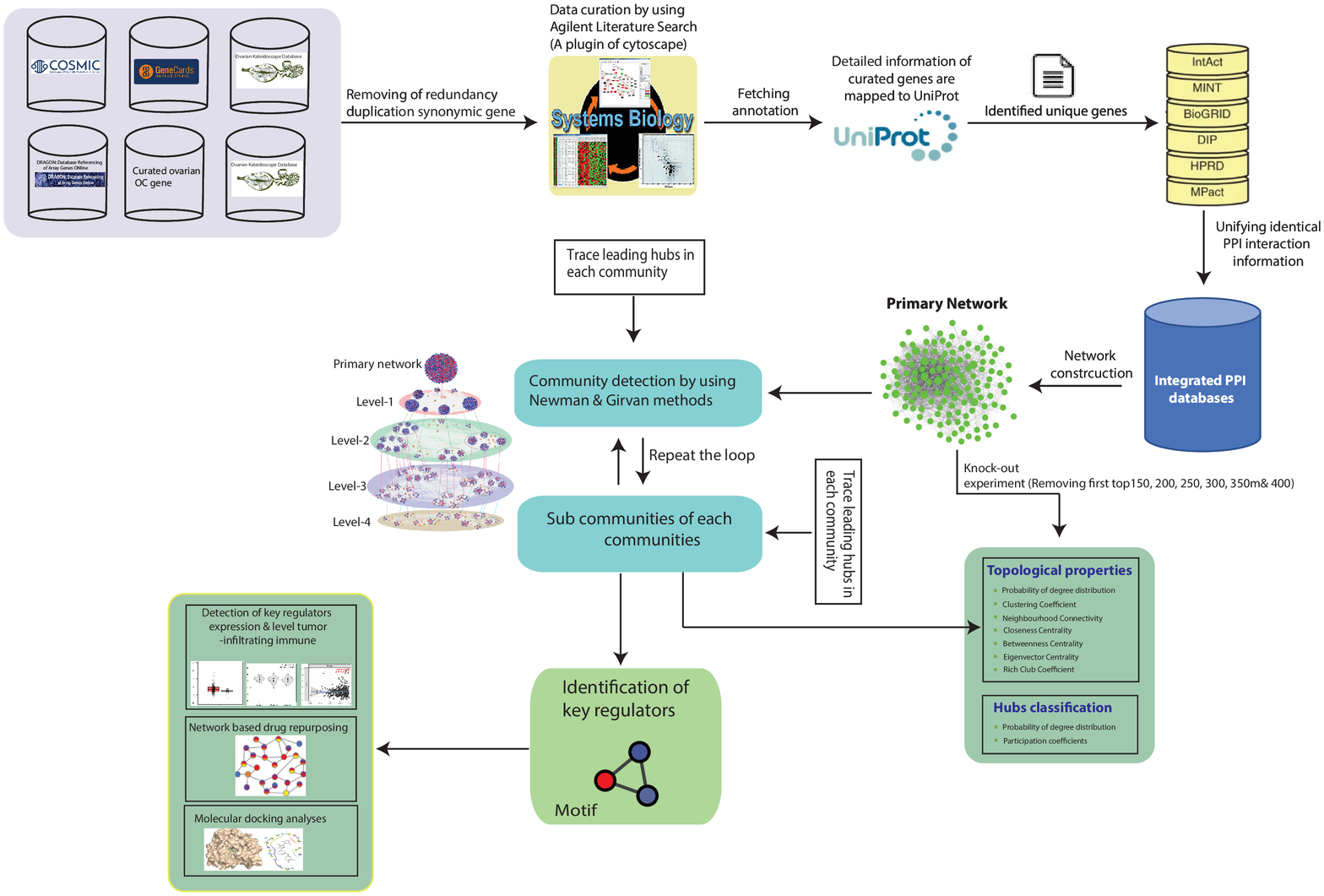}
\caption{Schematic diagram of the workflow of the methods implemented in the study of ovarian cancer network to drug discovery.} 
\end{center}
\end{figure}

\newpage

\begin{figure*}
\label{fig2}
\begin{center}
\includegraphics[height=15cm,width=16cm]{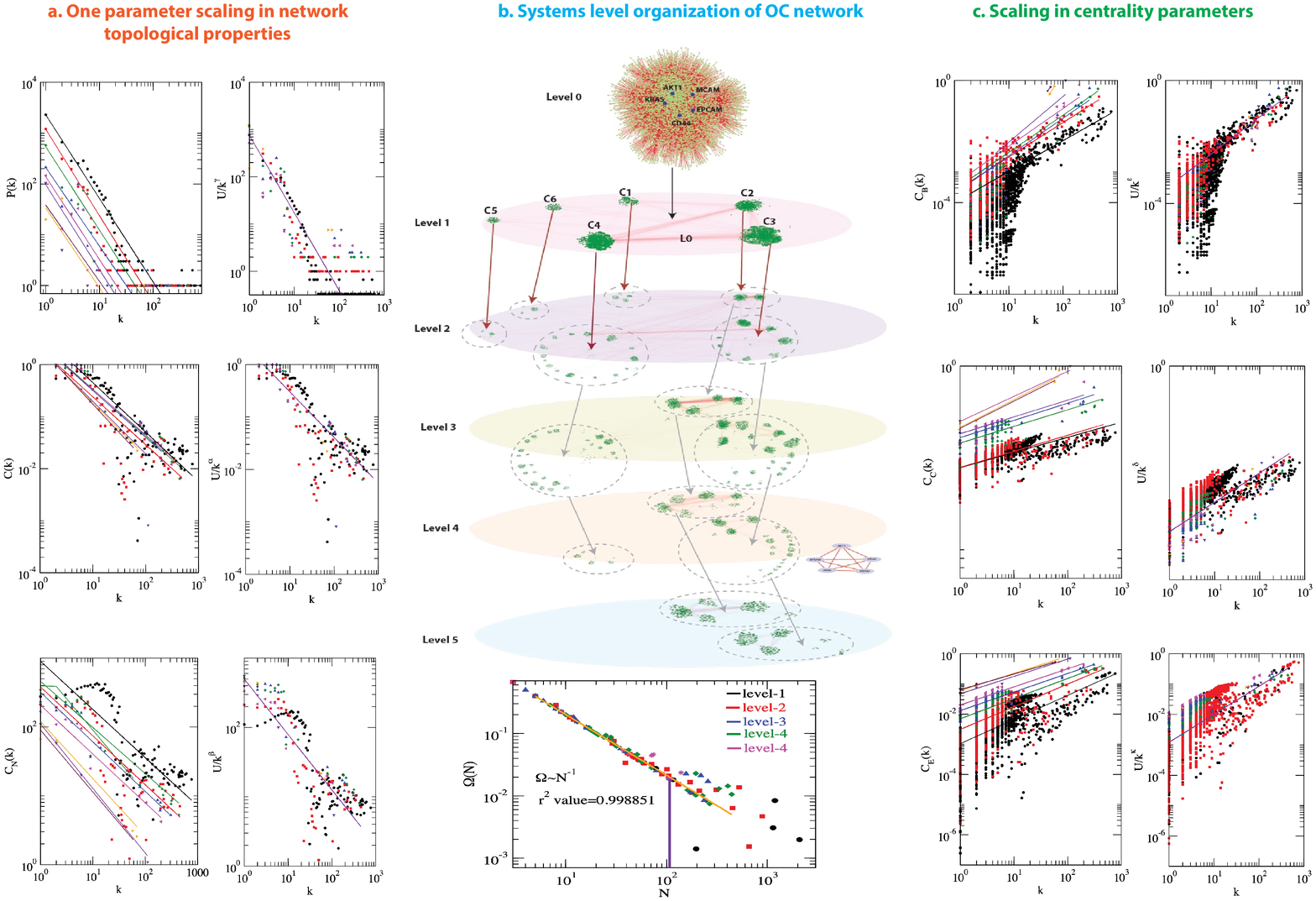}
\caption{\textbf{Properties of hierarchical organization of ovarian cancer network}: (a) Plots of $P(k), C(k)$ and $C_N(k)$ as a function of degree $k$ of networks, primary to motif level of the communities which incorporate novel key regulators and their corresponding collapse of the data points in a single curve following one parameter scaling law. The same is true for all the communities at various levels of organization which we did not show here. (b) Hierarchical organization of the complex OC network at various levels of organization. The calculated network connectance $\Omega$ as a function of network size $N$ showing one parameter scaling behavior. (c) Similar plots and behaviors of centrality parameters $C_B(k), C_C(k)$ and $C_E(k)$ indicating one parameter scaling nature.} 
\end{center}
\end{figure*}

\newpage

\begin{figure*}
\label{fig3}
\begin{center}
\includegraphics[width=1\textwidth]{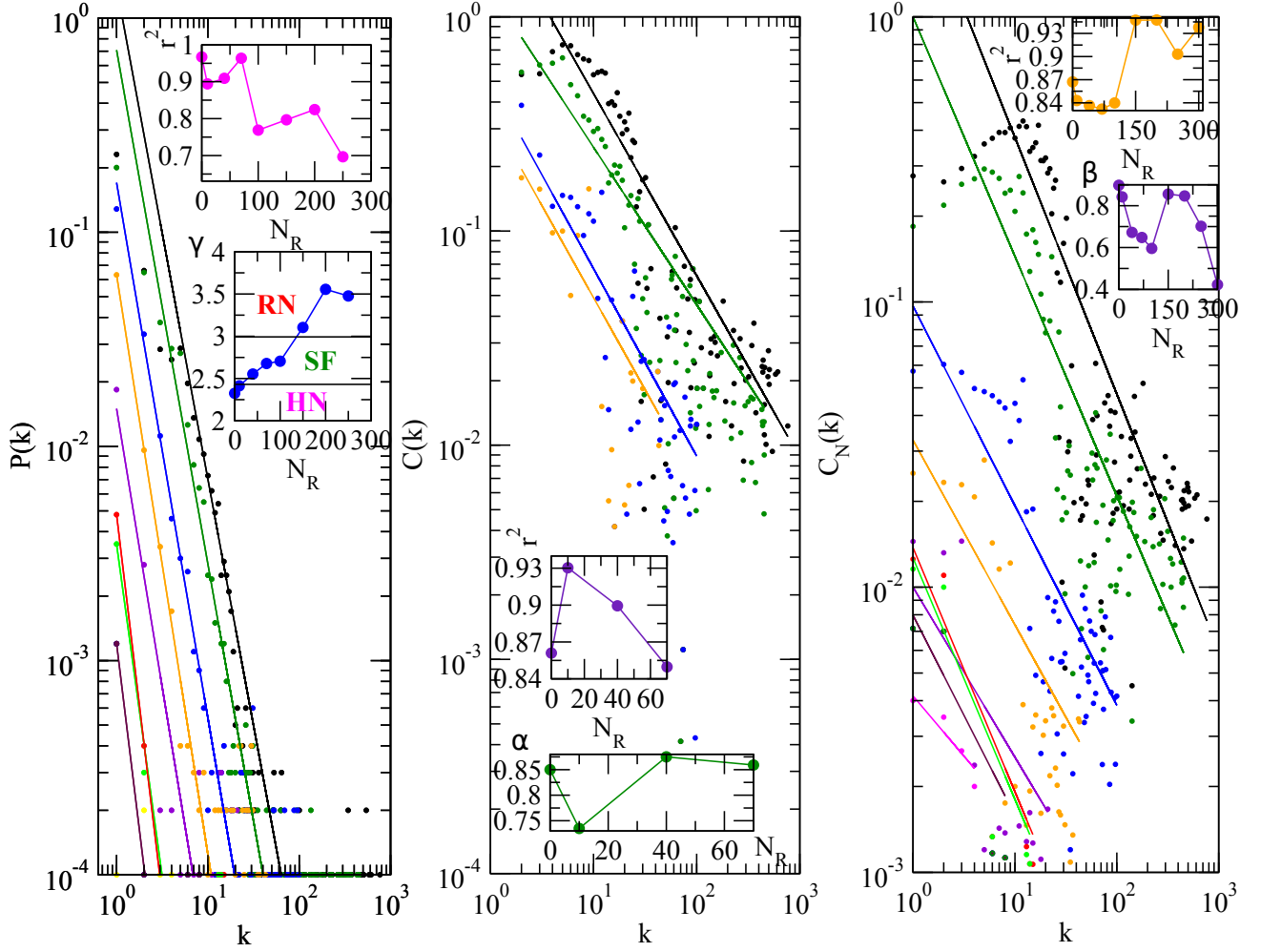}
\caption{\textbf{Hubs knock-out induced transition of network properties}: Topological properties of the systems level organization of ovarian cancer network after knock-out experiment by removing first top 150, 200, 250, 300, 350 and 400. The behaviors of degree distributions ($P(k)$), clustering co-efficient ($C(k)$), and neighbourhood connectivity ($C_N(k)$) after knock-out experiments. The insets show the changes in the exponent values by knock-out experiments of leading hubs and their statistical $r^2$ values of fitting procedure.} 
\end{center}
\end{figure*}

\newpage

\begin{figure}
\label{fig4}
\begin{center}
\includegraphics[height=13cm,width=18cm]{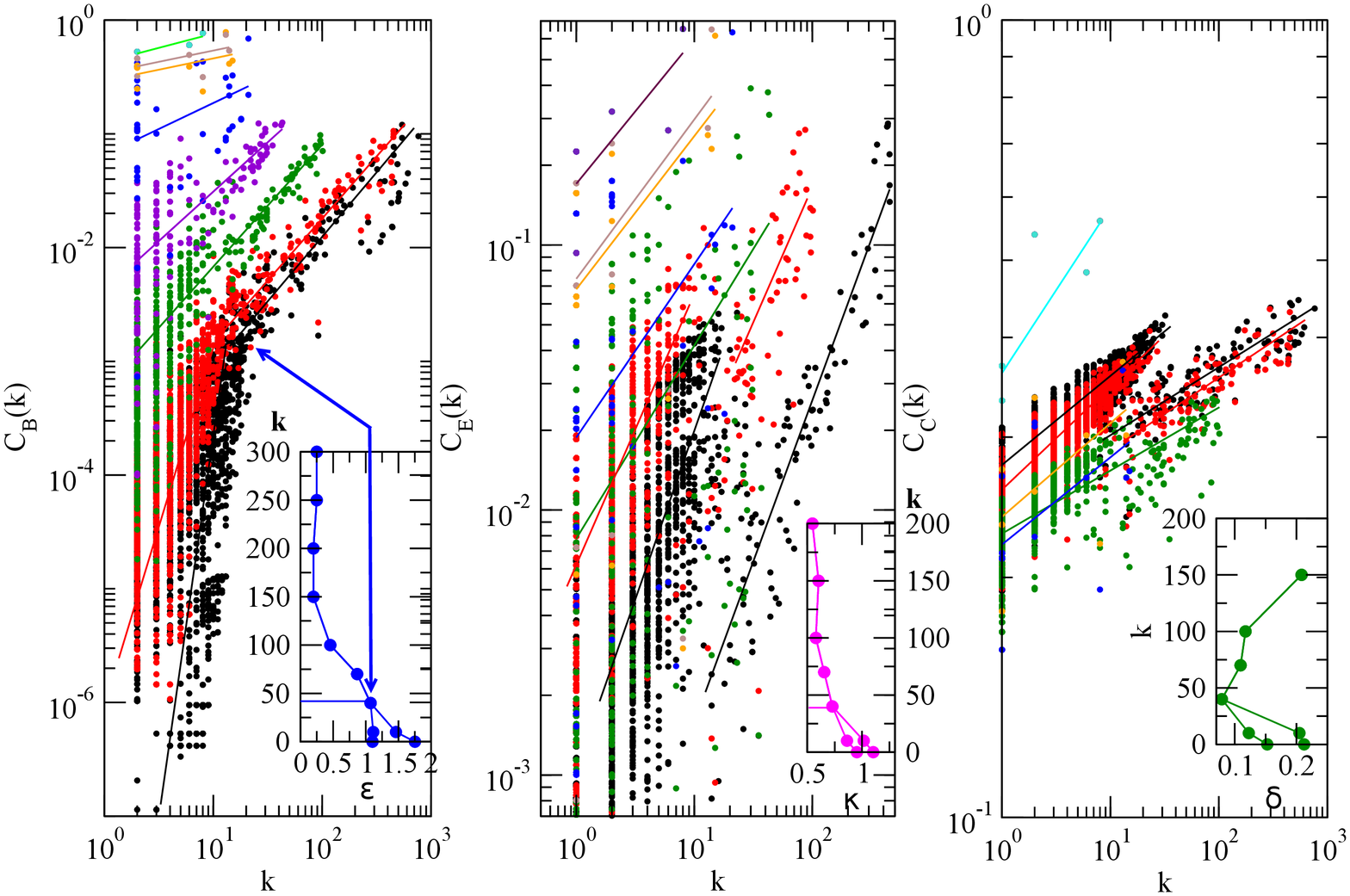}
\caption{\textbf{Hub knock-out induced properties of centrality parameters}: Properties of centrality measurements of the ovarian cancer network after knock-out experiments by removing first top 150, 200, 250, 300, 350 and 400. The behaviors of betweenness centrality ($C_B(k)$), closeness centrality ($C_C(k)$), and Eigenvector centrality ($C_E(k)$) after knock-out experiments. The insets show the changes in the exponent values by knock-out experiments of leading hubs and their statistical $r^2$ values of fitting procedure.} 
\end{center}
\end{figure}

\newpage

\begin{figure}
\label{fig5}
\begin{center}
\includegraphics[height=13cm,width=16cm]{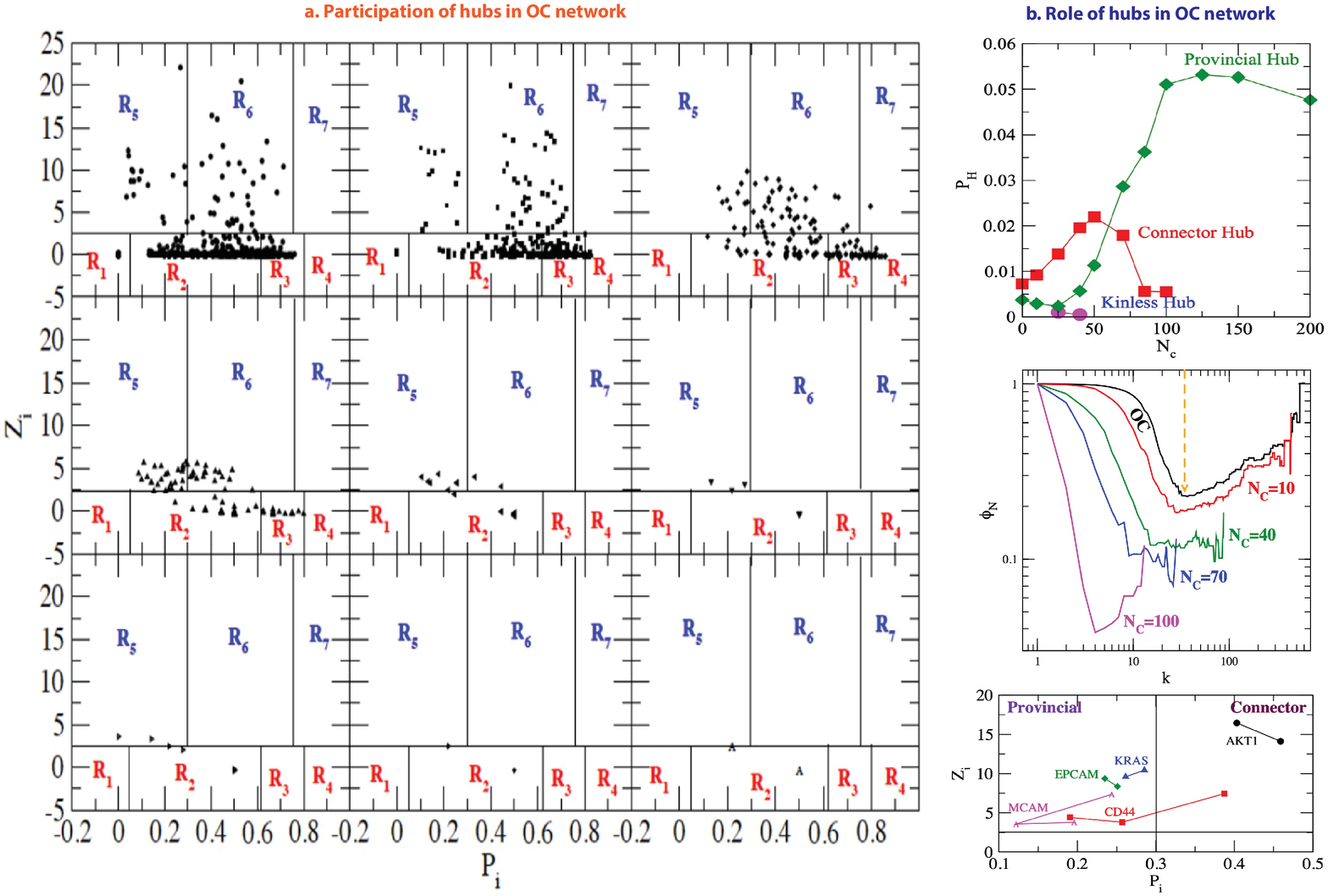}
\caption{\textbf{Involvement of different types of hubs in OC network regulation}: (a) Analysis of the role of hubs by analyzing participation ratio of the ovarian cancer network by plotting $Z_i$ as a function of $P_i$ in the knock-out experiments of leading hubs $N_C=10-200$.
(b) The plots indicating the probability of finding different types of hubs (Kinless, Connector and Provincial) as a function of $N_C$ (The number of leading hubs removed in the knock-out experiments). Rich-club analysis of ovarian cancer network, where, plots of $\phi$ as a function of degree $k$ are shown for knock-out experiments removing the first $N_C=10,40,70,100$ leading hubs from the primary ovarian cancer network. Location of key regulators of OC network in $(P_i,Z_i)$-plane.} 
\end{center}
\end{figure}

\newpage

\begin{figure}
\label{fig6}
\begin{center}
\includegraphics[height=14cm,width=17cm]{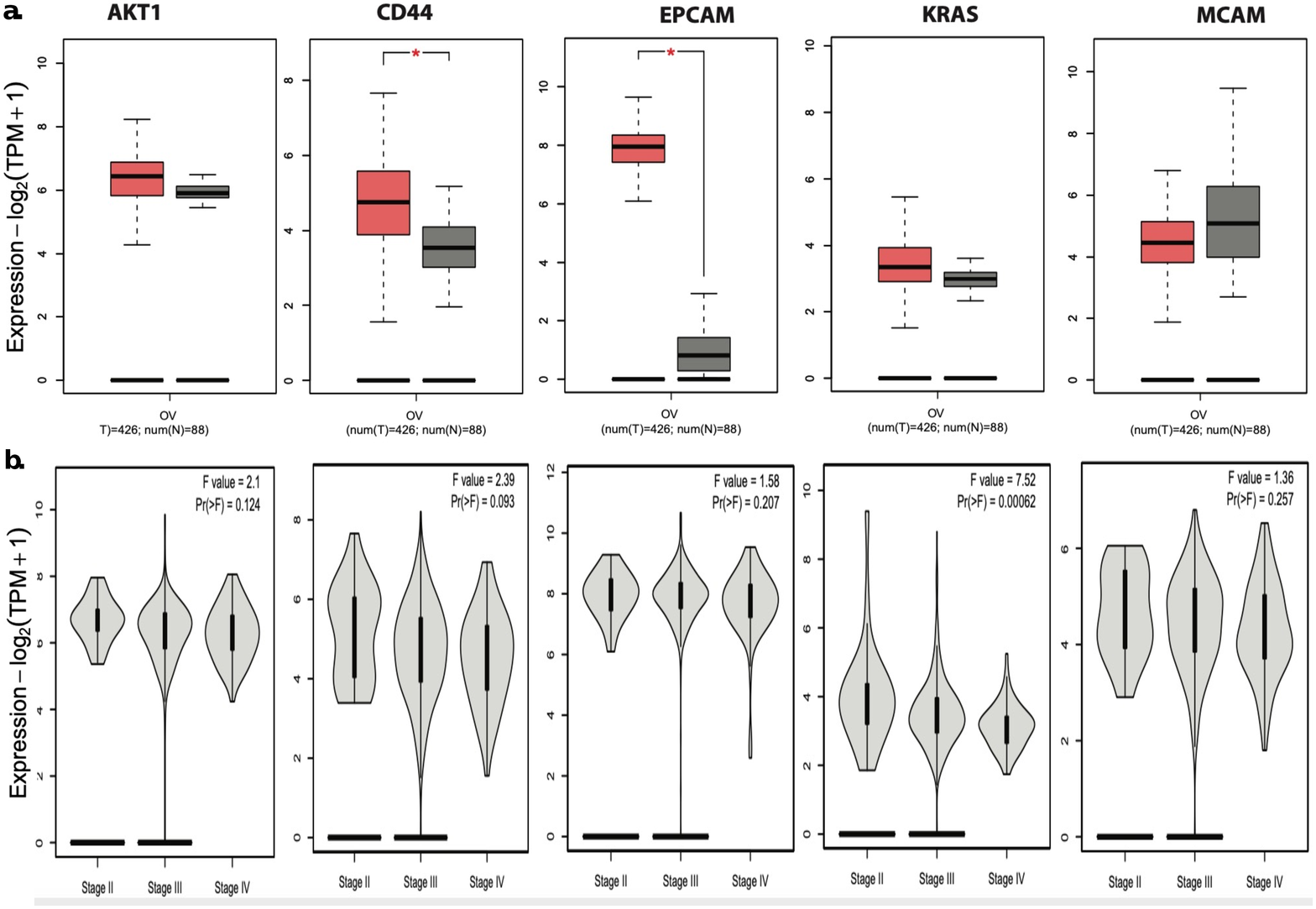}
\caption{\textbf{Gene expression patterns of key regulators using GEPIA database:} (a) Gene expression levels of key regulators (AKT1, CD44, EPCAM, KRAS and MCAM) of OC network which compare between normal (black color) OC patients (red color). (b) Plots of gene expression levels of the same key regulators of OC network at various cancer stages (I-IV).} 
\end{center}
\end{figure}

\newpage

\begin{figure}
\label{fig7}
\begin{center}
\includegraphics[height=17cm,width=17cm]{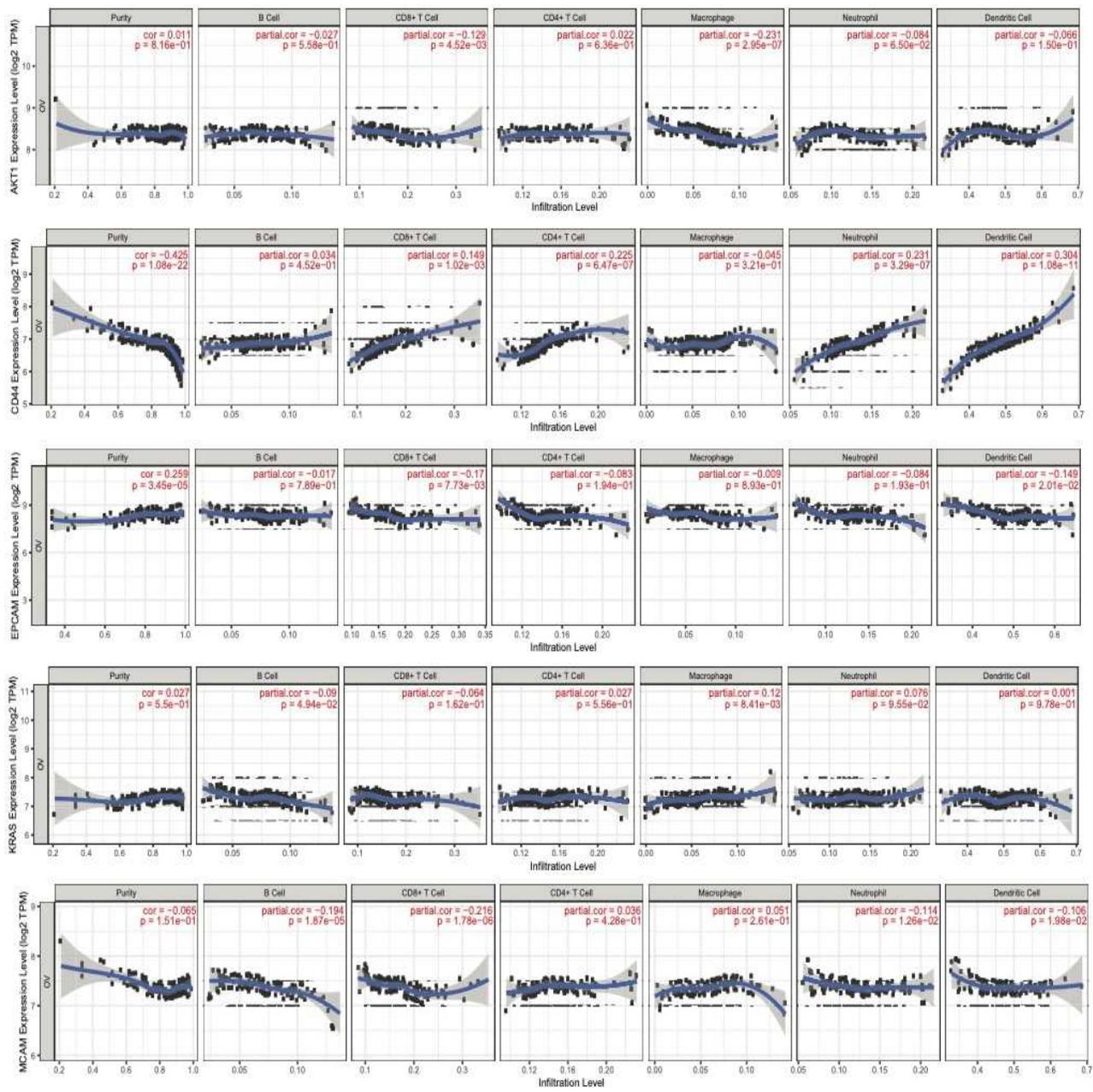}
\caption{\textbf{Correlation of the OC key regulators with various cancer-infiltrating immune subsets}: Plots of the gene expression levels of the OC key regulators (AKT1, CD44, EPCAM, KRAS and MCAM) as a function of infiltration levels for various cancer-infiltrating immune subsets via the TIMER database.} 
\end{center}
\end{figure}

\newpage

\begin{figure}
\label{fig8}
\begin{center}
\includegraphics[height=17cm,width=17cm]{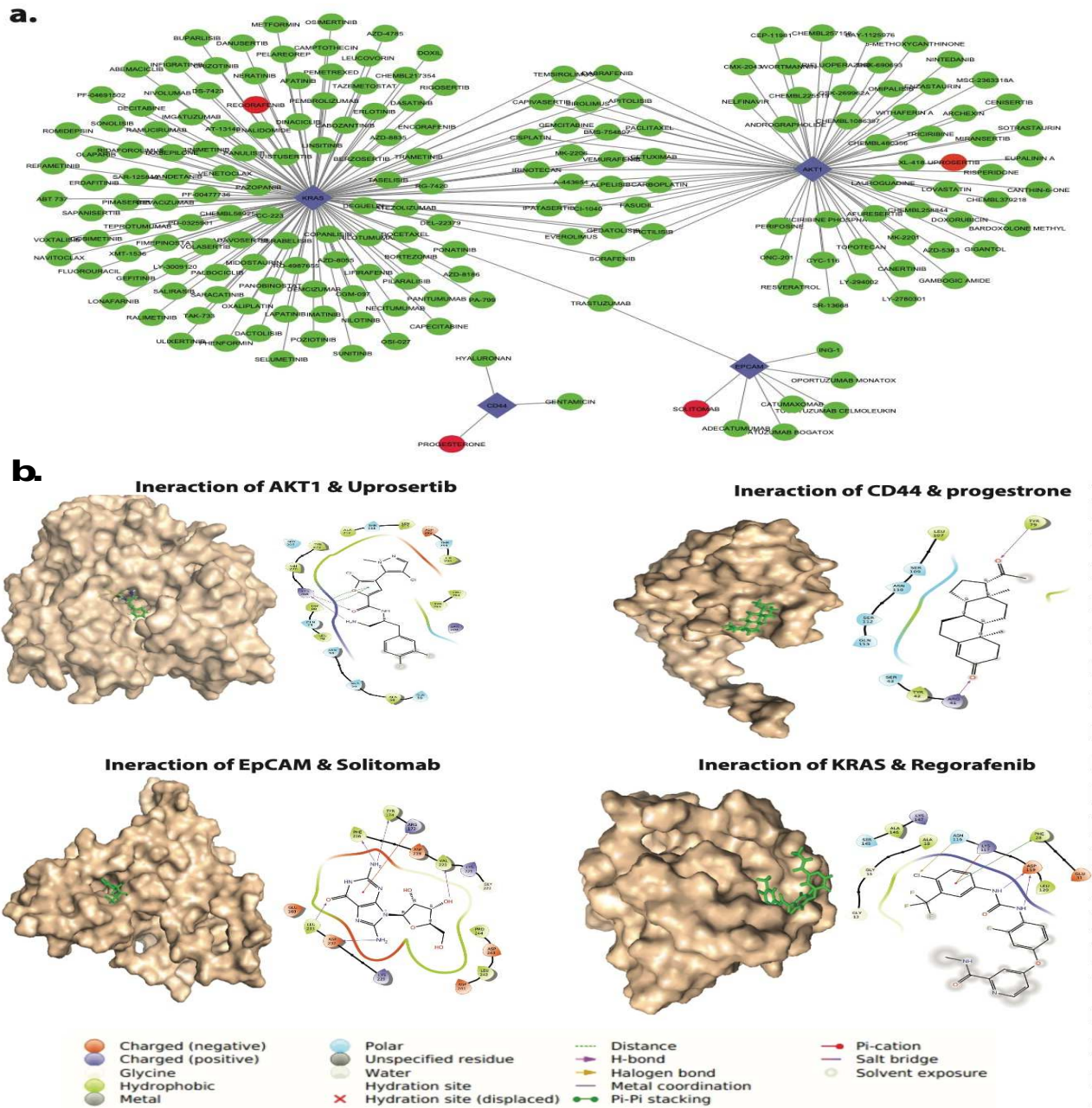}
\caption{\textbf{Identification of potential drug compounds which have optimal binding affinity with the key regulators}: (A) Interaction network of key regulators with FDA-approved drugs. The green circle represents drugs, blue color diamonds represents key regulators and red color drug shows highest score. (b) Molecular docking of key regulators and drugs with highest score. Uprosertib, Progesterone, Solitomab and Regorafenib drugs showed significant strong interactions with the key genes AKT1, CD44, EPCAM and KRAS respectively having the highest score.} 
\end{center}
\end{figure}

\end{document}